\documentclass{emulateapj}
\usepackage{apjfonts}
\usepackage{html}
\usepackage[squaren]{SIunits}
\submitted{Accepted for Publication in ApJ}

\def\gsim{\mathrel{\rlap{\lower2pt\hbox{\hskip0pt\small$\sim$}}
   \raise2pt\hbox{\small $>$}}}    
\def\lsim{\mathrel{\rlap{\lower2pt\hbox{\hskip0pt\small$\sim$}}
   \raise2pt\hbox{\small $<$}}}    
\def\bq{\begin{equation}} 
\def\eq{\end{equation}}

\newcommand{\bqa}{\begin{eqnarray}} 
\newcommand{\eqa}{\end{eqnarray}}

\newcommand{\rperi}{r_{\rm peri}}
\newcommand{\rapo}{r_{\rm apo}}
\newcommand{\rcirc}{r_{\rm circ}}

\newcommand{\vmx}{v_{\rm max}}
\newcommand{\rmx}{r_{\rm max}}
\newcommand{\pmx}{\rho_{\rm max}}

\newcommand{\smjustify}{\hspace{0.15cm}}

\def\smhwidth{0.48\textwidth}

\def\bulletskip{0.15cm}

\addunit{\mass}{M_\odot}
\addunit{\Mpc}{Mpc}
\addunit{\kpc}{kpc}
\addunit{\speed}{km\ s^{-1}}

\shorttitle{Evolution of Low Mass Galactic Subhalos and Dependence on Concentration}

\begin{document}

\title{Evolution of Low Mass Galactic Subhalos and Dependence on Concentration}

\author{J.~D.~Emberson\altaffilmark{1,2}, Takeshi~Kobayashi\altaffilmark{1,3} \& Marcelo~A.~Alvarez\altaffilmark{1}}
\altaffiltext{1}{Canadian Institute for Theoretical Astrophysics,
  University of Toronto, 60 St.George St., Toronto, ON M5S 3H8, Canada} 
\altaffiltext{2}{Department of Astronomy and Astrophysics, University
  of Toronto, 50 St. George, Toronto, ON M5S 3H4, Canada} 
\altaffiltext{3}{Perimeter Institute for Theoretical Physics, 31 Caroline Street North, Waterloo, ON N2L 2Y5, Canada}
\email{emberson@astro.utoronto.ca}

\begin{abstract}
We carry out a detailed study of the orbital dynamics and structural
evolution of over 6000 subhalos in the Via Lactea II simulation, from
infall to present.  By analyzing subhalos with masses down to
$m = \unit{4\times10^5}{\mass}$, we find that lower mass subhalos, which
are not strongly affected by dynamical friction, exhibit behaviors qualitatively
different from those found previously for more massive ones. 
Furthermore, there is a clear trend of subhalos that fell into
the host earlier being less concentrated. We show that the
concentration at infall characterizes various aspects of  subhalo
evolution. In particular, tidal effects truncate the
growth of less concentrated subhalos at larger distances from the
host; subhalos with smaller concentrations have larger infall
radii. The concentration at infall is further shown to be a
determining factor for the subsequent mass loss of subhalos within
the host, and also for the evolution of their internal structure
in the $v_{\mathrm{max}} - r_{\mathrm{max}}$ plane. Our findings raise the prospects of using the concentration to predict the tidal evolution of subhalos, which will be useful for obtaining analytic models of galaxy formation, as well as for near field cosmology.
\end{abstract}


\section{Introduction}

The standard model of cosmological structure formation is based upon the notion that the gravitational landscape is dominated by cold dark matter (CDM), which initially collapses on small scales and grows hierarchically to larger scales through the continued merger and accretion of smaller objects. Many of the accreting systems survive to the present epoch as independent entities within their host, giving rise to a system of nested substructure within the largest objects to have formed today. These surviving remnants present a unique opportunity to study the seeds of galaxy formation and the chance to probe the nature of dark matter on small scales. 

Accomplishing these goals requires the development of an accurate and predictive theory for the  evolution of substructure. This remains a difficult task and a fundamental problem in the burgeoning area of near field cosmology. Even without taking into account the complicating role played by dissipative effects associated with, e.g., star formation and feedback, one must still contend with reconciling the stochasticity in the primordial fluctuations that seed the subsructure in any given object, on the one hand,  and the highly nonlinear gravitational dynamics associated with tidal disruption and dynamical friction, on the other hand. As is often the case, efforts to tackle this problem generally fall into one of two categories. 

First are direct numerical simulations \citep[e.g.,][]{diemand/etal:2008a,springel/etal:2008,garrison-kimmel/etal:2014}, which attempt to solve the problem {\em ab initio} from cosmological initial conditions zoomed on a single host. This class of approaches suffers from small number statistics, both in the number of individual objects simulated and in the range of underlying background cosmologies (warm dark matter, self-interacting dark matter, broken scale invariance, etc.). In addition, while simulations have begun to converge on accurate solutions for individual systems in a $\Lambda$CDM universe composed only of collisionless dark matter, star formation and feedback must still be treated using heuristic sub-grid approaches calibrated with empirical data, blurring the line between theory and observation and complicating the interpretation of simulations.

The second class involves semi-analytical galaxy formation models \citep[e.g.][]{taylor/babul:2004a,zentner/etal:2005,gan/etal:2010,jiang/vandenbosch:2014,pullen/etal:2014} which have a long history in cosmology, and have begun to be successfully applied to the local universe. The standard approach is to generate a mass accretion history using the excursion set formalism \citep[e.g.,][]{bond/etal:1991,lacey/cole:1993} followed by an integration of individual accreting orbits from the moment of infall to the present day. Orbital parameters at infall are drawn from probability distributions motivated by numerical simulations \citep[e.g.,][]{navarro/etal:1995,tormen/etal:1997,ghigna/etal:1998,benson:2005,zentner/etal:2005,wang/etal:2005,khochfar/burkert:2006,jiang/etal:2008,wetzel:2011,jiang/etal:2014b} and the time integration contains prescriptions for various nonlinear processes such as tidal stripping, tidal heating, and dynamical friction. The utility of this approach is its computational speed, allowing one to simulate multiple realizations and cover a broad region of model and parameter space compared to what can be achieved using expensive cosmological simulations. The drawback is that any simplifying assumptions (e.g., symmetries in the host potential, omission of substructure interaction) inherent to the model may affect the final result in an unknown or unphysical manner. 

In this paper, we present a detailed case study of an individual object simulated at high resolution -- the Via Lactea II \citep[][VL2]{diemand/etal:2008a} simulation -- with the aim of making connections relevant to semi-analytic models of substructure evolution. We focus on a self-consistent description of the most important physical processes and relationships, rather than on direct comparison to specific observations. Our goal is to separate the robust quantitative predictions of this simulation from those that are unique to the particular background cosmology and random realization used to generate its initial conditions. 

This paper is organized as follows. In Section \ref{sec:vl2data} we describe our methodology of extracting substructure evolution from the public VL2 catalogues. In Section \ref{sec:results} we present the main results of our work. We begin in \S\ref{sec:results1} with a basic description of the host halo and in \S\ref{sec:results2} statistics of its subhalo population, followed in \S\ref{sec:results3} with a presentation of orbital properties at the time of infall, and in \S\ref{sec:results4} with a quantitative assessment of substructure evolution including the physical processes of tidal mass loss and its dependence on subhalo properties, the orbital timescale, changes in the orientation of the orbital plane, and the dynamical readjustment of the internal structure of subhalos as portrayed by their movement in the $r_{\rm max}-v_{\rm max}$ plane. In \S\ref{sec:dispop} we compare the surviving and disrupted subhalo populations of VL2 and investigate how survivability depends on infall redshift, mass, concentration, and orbital parameters. We summarize our conclusions in Section \ref{sec:summary}.


\section{Data Analysis}
\label{sec:vl2data}

The VL2 simulation traced the growth of a galactic host halo within a high-resolution region sampled with roughly one billion particles of mass \unit{4100}{\mass}. In what follows we make use of the main halo catalogue made publicly available\footnote{\url{http://www.ucolick.org/~diemand/vl/data.html}} by the VL2 team. This catalogue contains evolutionary tracks of all 20048 (sub)halos within the simulation box that are resolved at $z = 0$ and for which their peak circular velocity was larger than $\vmx = \unit{4}{\speed}$ at some time during their evolution. The latter restriction is imposed to discard small halos affected by insufficient resolution. 

The catalogue contains a collection of halo properties at 27 discrete redshifts between $0 \leq z \leq 27$. These properties include: the $x$, $y$, and $z$ positions and velocities relative to the host halo rest frame; the tidal radius, $r_{\rm tid}$, and tidal mass, $m_{\rm tid}$; the maximum of the circular velocity curve, $v_{\rm max}$, and the radius, $r_{\rm max}$, at which this occurs. Empty values occur at redshifts when the halo progenitor either did not exist or overlapped with a more massive halo. In what follows we consider only the redshift range $0 \leq z \leq 4.56$ for which the host progenitor was consistently identified within the simulation. This contains 19 redshift snapshots which we further refine by performing cubic spline interpolations of the above quantities to generate a total of 181 discrete sample points equally spaced by 68.8 Myr.

Subhalos are identified in VL2 using the six-dimensional phase-space friends-of-friends (6DFOF) algorithm described in \citet{diemand/etal:2006}. Around each (sub)halo the circular velocity profile, $v_{\rm circ} = \sqrt{G m(<r)/r}$, is computed in spherical bins and is fitted with the sum of contributions from an NFW profile and a constant density background, $\rho_{\rm bg}$. The latter component is then subtracted from the (sub)halo density profile and a tidal radius is computed by solving $\rho_{\rm sub}(r_{\rm tid}) = 2\rho_{\rm bg}$, corresponding to the tidal radius of an isothermal sphere on a circular orbit within an isothermal host \citep{diemand/etal:2007a}. The tidal mass is assigned $m_{\rm tid} = m(<r_{\rm tid})$. For sufficiently isolated halos, where the background density is small and $r_{\rm tid} > r_{200}$ (the radius at which the enclosed density is 200 times the mean {\em matter} density), $r_{\rm tid}$ is capped at $r_{200}$ and $m_{\rm tid} = m_{200}$ (Diemand, private communication).

Hence, the subhalo masses used in this paper are not the result of an unbinding procedure of dark matter particles. Nevertheless, it was shown in the Via Lactea I (VL1) analysis \citep{diemand/etal:2007} that this definition of tidal mass indeed agrees well with the true bound mass when the subhalo is near apocenter, but may significantly underestimate bound mass near pericenter. For this reason we generally only report mass quantities near apocenter and explicitly point out to the reader when this is not the case. 

In the following subsections we define concepts and present our methodology of investigating substructure evolution from the VL2 data. We begin in \S\ref{subsec:subdef} with the definition of a subhalo. In \S\ref{subsec:vl2con} we model the internal structure of the host and its subhalos via concentration parameters. We define in \S\ref{subsec:infdef} the redshift, $z_{\rm infall}$, at which a subhalo is said to first infall onto the host. In \S\ref{subsec:eldef} we outline our calculations of orbital energy and angular momentum and finish in \S\ref{subsec:orbdef} with a description of our method of tracing subhalo orbits after infall. 

As a matter of convenience, we remove explicit redshift dependence in our following notation and remind the reader here that all quantities are computed at discrete times. We use lower case notation (e.g., $m$, $r_{\rm max}$, $v_{\rm max}$) when referring to subhalos while upper case notation (e.g., $M$, $R_{\rm max}$, $V_{\rm max}$) is reserved for the host. The mass of a subhalo is taken to be its tidal mass while the mass of the host is taken to be its virial mass (see \S\ref{subsec:vl2con}). We often use $\mu \equiv m(z)/M(z)$ to denote the instantaneous mass ratio between a subhalo and the host. At times we normalize to the present-day host mass in which case we define $\mu_0 \equiv m(z)/M(0)$. In what follows we assume the same cosmology as the VL2 simulation; namely, the  $\Lambda$CDM parameters ($\Omega_m$, $\Omega_\Lambda$, $h$, $n_s$, $\sigma_8$) = (0.238, 0.762, 0.73, 0.951, 0.74) from the WMAP 3-year data release \citep{spergel/etal:2007}.

\subsection{Definition of a subhalo}
\label{subsec:subdef}

We flag an object in the VL2 catalogue as a subhalo if at one time during its evolution it passed within the instantaneous virial radius of the host. This definition includes subhalos that are presently {\em within} the virial radius as well as subhalos that currently reside {\em outside} the virial radius. We refer to the latter group as {\em ejected} subhalos in the sense that they are now removed from the virial boundary of the host. This terminology, however, does not imply that these subhalos are unbound from the host, as shown in \S\ref{sec:results-etainf}.

In later sections we explore subhalo tidal mass loss. It was shown in \citet{kazantzidis/etal:2004} that subhalos with too few particles within their tidal radius experience artificially large tidal mass loss. For this reason we impose a further restriction on the VL2 catalogue that only objects with at least 100 particles in their tidal radius at $z = 0$ may be considered as subhalos. This sets a minimum mass resolution of $m = \unit{4\times10^5}{\mass}$. 

We find a total of 7569 objects meeting the above criteria. 5845 (77\%) of these currently reside within the host virial radius of $R_{\rm vir} = \unit{320}{\kpc}$ (see \S\ref{subsec:vl2con}) while the remaining 1724 (23\%) are currently ejected. For the remainder of the paper we exclude those subhalos whose infall (see \S\ref{subsec:infdef}) is determined to be $z_{\rm infall} > 4.56$. This reduces the total population to 6145 subhalos with 4607 (75\%) currently within the virial radius and 1538 (25\%) ejected. 

\subsection{Host and subhalo mass distributions}
\label{subsec:vl2con}

It was shown in \citet{navarro/etal:1997} that dark matter halos obey a universal density profile, named an NFW profile after its founders. This has the form $\rho(r) \propto x^{-1}(1+x)^{-2}$, where $x\equiv r/r_s$ and $r_s$ is the radius at which $d\ln\rho/d\ln{r}=-2$. The virial radius, $r_{\rm vir}$, is defined such that the enclosed density is $\Delta(z)$ times the critical density, where $\Delta(z)$ is calculated using the fitting function to the overdensity of a virialized uniform sphere in a flat universe given in \citet{byran/norman:1998}. An NFW profile is often parameterized by its concentration, $c_{\rm vir} \equiv r_{\rm vir}/r_s$, which describes the degree to which the mass is contained within the central region.

We assume that the density profile of the host follows an NFW form. We determine its concentration by finding the unique NFW profile for which the mass enclosed within $R_{\rm max}$ is $R_{\rm max}V_{\rm max}^2/G$. This involves the implicit solution of
\bq
g(C_{\rm vir})=g(x_m)\frac{\Delta(z)}{2}\left[\frac{H(z)R_{\rm max}}{V_{\rm max}}\right]^2,
\label{eq:croot}
\eq
where $g(x) \equiv f(x)/x^3$ with $f(x) \equiv \ln(1+x)-x/(1+x)$, $H(z)$ is the Hubble parameter, and $x_m\equiv R_{\rm max}/r_s\approx 2.163$. Once the concentration is obtained, the host halo mass is computed as $M_{\rm vir}=4\pi\rho_{\rm crit}{\Delta}R_{\rm vir}^3/3$ where $R_{\rm vir}=C_{\rm vir}R_{\rm max}/x_m$.

The assumption of an NFW profile for the host should be valid over the entire redshift range considered here. For subhalos, however, an NFW profile is only valid up until its moment of infall onto the host. It was shown by \citet{hayashi/etal:2003} that the processes of tidal heating and stripping tend to modify the internal structure of subhalos away from their initial form. For this reason, we only use $c_{\rm vir}$ obtained from equation (\ref{eq:croot}) for subhalos at their time of infall. Afterwards, we define a concentration parameter
\bq
c_{\rm max} = 2\left[ \frac{v_{\rm max}}{H_0 r_{\rm max}} \right]^2,
\label{eq:cproxy}
\eq
which gives the mean density within $r_{\rm max}$ in units of the critical density. Comparing to equation (\ref{eq:croot}) shows that, for any given redshift, there exists a monotonic relationship between $c_{\rm vir}$ and $c_{\rm max}$.

\subsection{Definition of infall}
\label{subsec:infdef}

In an idealized description, a subhalo will form distinct from its future host, accreting surrounding material and growing steadily in size. This process will occur until the time at which tidal interactions with its host become important. At this point, the combined action of dynamical friction and tidal stripping will cause the subhalo to lose mass over time. We therefore define infall as this turnaround phase in the growth history of the subhalo. That is, we define the redshift, $z_{\rm infall}$, of infall onto the host to be the moment in time at which the mass of the subhalo is a maximum\footnote{This definition does not filter out the possibility that a subhalo may initially lose mass via tidal interactions with a halo other than its future host. Such group preprocessing was studied in \citet{wetzel/etal:2015} where it was found that a significant fraction of subhalos reside within the virial radius of a different halo prior to passing through the virial radius of the main host. We therefore note the possibility that our  $z_{\rm infall}$ are biased toward larger values, though it is unclear to what magnitude group preprocessing affects premature mass loss.}. 

Recall that masses in VL2 are assigned as the mass contained within the tidal radius. The tidal radius is derived, at any moment, by equating the subhalo density profile to twice the local background density. For sufficiently isolated subhalos, where the background density is small, the tidal radius is capped at $r_{200}$. The resultant tidal mass provides a good estimation of the true bound mass when the subhalo is near apocentre, which is generally the case at infall. As mentioned earlier, we consider only the 6145 subhalos for which $z_{\rm infall} \leq 4.56$ since at earlier times the host progenitor is only sporadically identified within the VL2 catalogues, preventing us from computing orbital properties at those times. 

An alternative convention that is commonly used in the literature is to define infall as the moment the subhalo passes through the virial radius of the host. However, as shown previously \citep{hahn/etal:2009,behroozi/etal:2014}, subhalos generally undergo strong tidal forces at distances larger than $R_{\rm vir}$. Furthermore, the virial radius evolves with redshift through its dependence on $\Delta(z)$, meaning that its value will change even if the intrinsic mass profile of the host is unchanging. The virial radius is therefore not well-suited for defining the distance at which a subhalo becomes tidally truncated by the host, and can be said to undergo infall in the sense considered here.

\subsection{Orbital energy and angular momentum}
\label{subsec:eldef}

We determine the energy and angular momentum by assuming that subhalos evolve as isolated point particles within the spherically symmetric NFW profile of the host. In this case, the host potential is  
\bq
\Phi(r) = - \frac{R_{\rm max} V_{\rm max}^2}{f(x_m)} \frac{{\rm ln}(1+ x_m r/R_{\rm max})}{r},
\label{eq:nfwpot}
\eq
where $r = |{\bf r}|$ is the radial separation between the subhalo and host. In this expression we have taken the zero point of the potential to be at infinity. The specific orbital energy of the subhalo is
\bq
E = \frac{1}{2} {\bf v}\cdot{\bf v} + \Phi(r),
\label{eq:orben}
\eq
while its specific orbital angular momentum is
\bq
L = {\bf r} \times {\bf v}.
\label{eq:orbang}
\eq
Here ${\bf v}$ is the physical relative velocity between the subhalo and host which includes the sum of peculiar motion and Hubble flow.

In general, the continued action of dynamical friction will steadily drain energy and angular momentum from subhalo orbits. The evolution of a subhalo after infall thus depends strongly on its initial energy and angular momentum. It is therefore important to characterize the infall distributions of $E$ and $L$ as inputs in semi-analytic models of subhalo evolution. A common convention is to normalize these quantities in terms of a circular orbit of the same energy. We introduce two variables: (i) $\eta \equiv \rcirc/R_{\rm vir}$, defined to be the ratio of the radius, $\rcirc$, of a circular orbit of the same energy $E$ as the subhalo to the virial radius, $R_{\rm vir}$, of the host at infall; (ii) the circularity, $\epsilon \equiv L / L_{\rm circ}$, defined to be the ratio of the subhalo angular momentum, $L$, to the angular momentum, $L_{\rm circ}$, of a circular orbit of the same energy. To compute $\eta$ and $\epsilon$ we must first evaluate $\rcirc$, which is achieved by numerically solving the expression
\bq
\frac{{\rm ln}(1+y)}{y} + \frac{1}{1+y} = - \frac{2 E f(x_m)}{x_m V_{\rm max}^2},
\label{eq:orbrcirc}
\eq
where $y \equiv x_m \rcirc / R_{\rm max}$. Then $L_{\rm circ} = \sqrt{G M(r_{\rm circ}) r_{\rm circ}}$ where $M(r_{\rm circ})$ is the mass contained within radius $r_{\rm circ}$ of the host.

\begin{figure} 
\begin{center}
\includegraphics[width=\smhwidth]{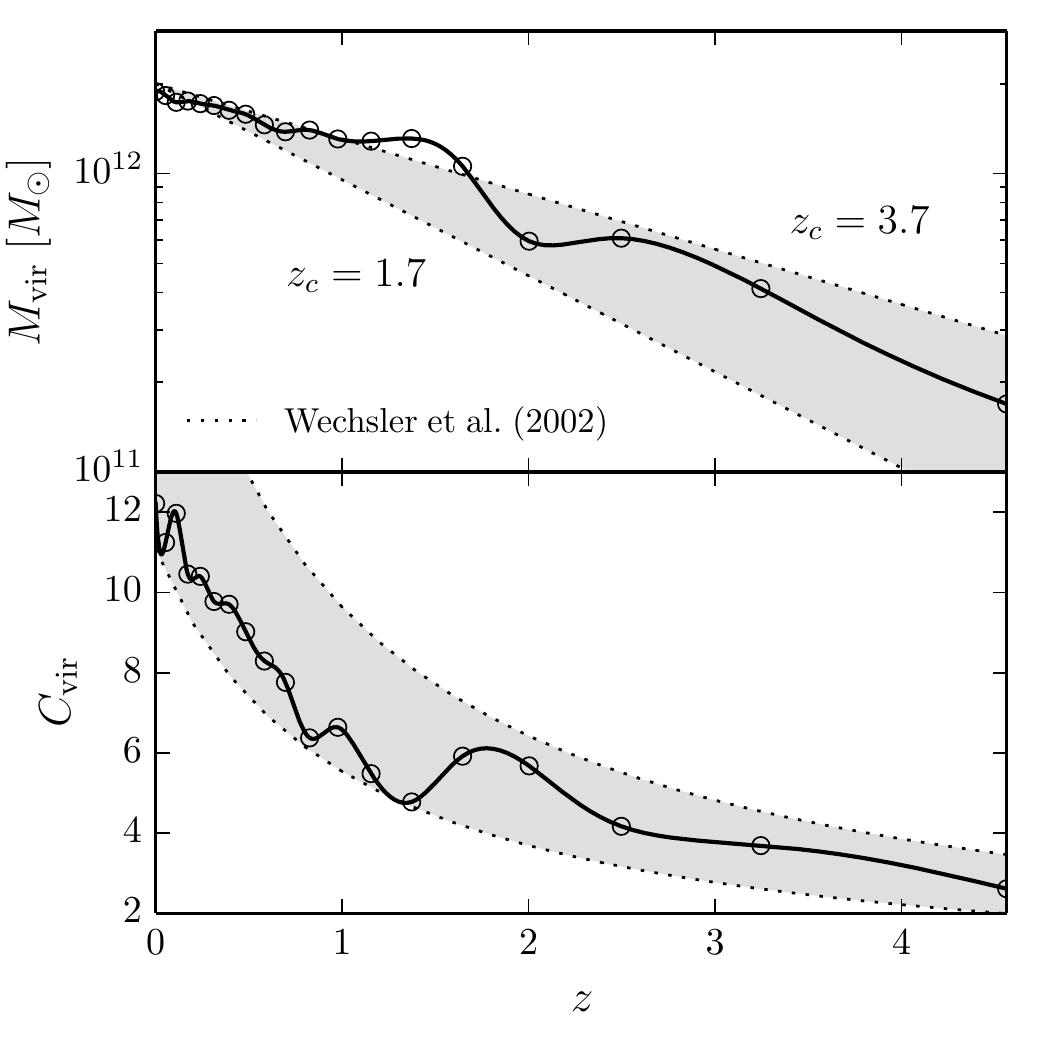}
\vspace{-0.8cm}
\end{center}
\caption{Evolution of the viral mass (top panel) and concentration (bottom panel) of the host halo obtained by finding the unique NFW profile that matches the values of $R_{\rm max}$ and $V_{\rm max}$ at each redshift. Open circles denote redshifts at which the VL2 catalogues are sampled while solid black lines trace the results derived from a cubic spline interpolation of $R_{\rm max}$ and $V_{\rm max}$. The dotted black lines in each panel show the expected evolution for the fitting functions given by \citet{wechsler/etal:2002} for collapse times $z_c = 1.7$ and $z_c = 3.7$ (see text). The VL2 curves are well contained within the shaded region, which may reflect an initial ``collapse'' at $z \sim 3.7$ (in the sense of the Wechsler et al. formalism) followed by an episode of significant mass accretion at $z \sim 1.7$ which ``resets'' the concentration back to the virialization value of $C_{\rm vir} \sim 4$.}
\label{fig:mwh} \vspace{0.2cm}
\end{figure}

The definitions of $\eta$ and $\epsilon$ used here are self-consistent with the description of a subhalo orbiting within an isolated NFW profile. This does not, however, conform with the standard method applied in semi-analytic models of substructure evolution. Instead, it is common to report these quantities at the time when the subhalo first crosses through $R_{\rm vir}$ and to model the host potential as a point mass of $M_{\rm vir}$. In this case, the orbital energy is
\bq
E = \frac{1}{2} {\bf v}\cdot{\bf v} - \frac{GM_{\rm vir}}{R_{\rm vir}},
\label{eq:orbeng-sa}
\eq
and the radius of a circular orbit of the same energy is
\bq
r_{\rm circ} = -\frac{G M_{\rm vir}}{2 E}.
\label{eq:orbcirc-sa}
\eq
When discussing $\eta$ and $\epsilon$ at infall we report the results of both methods so that we can make direct comparisons to previous work. 

\begin{figure*} 
\begin{center}
\includegraphics[width=0.9\textwidth]{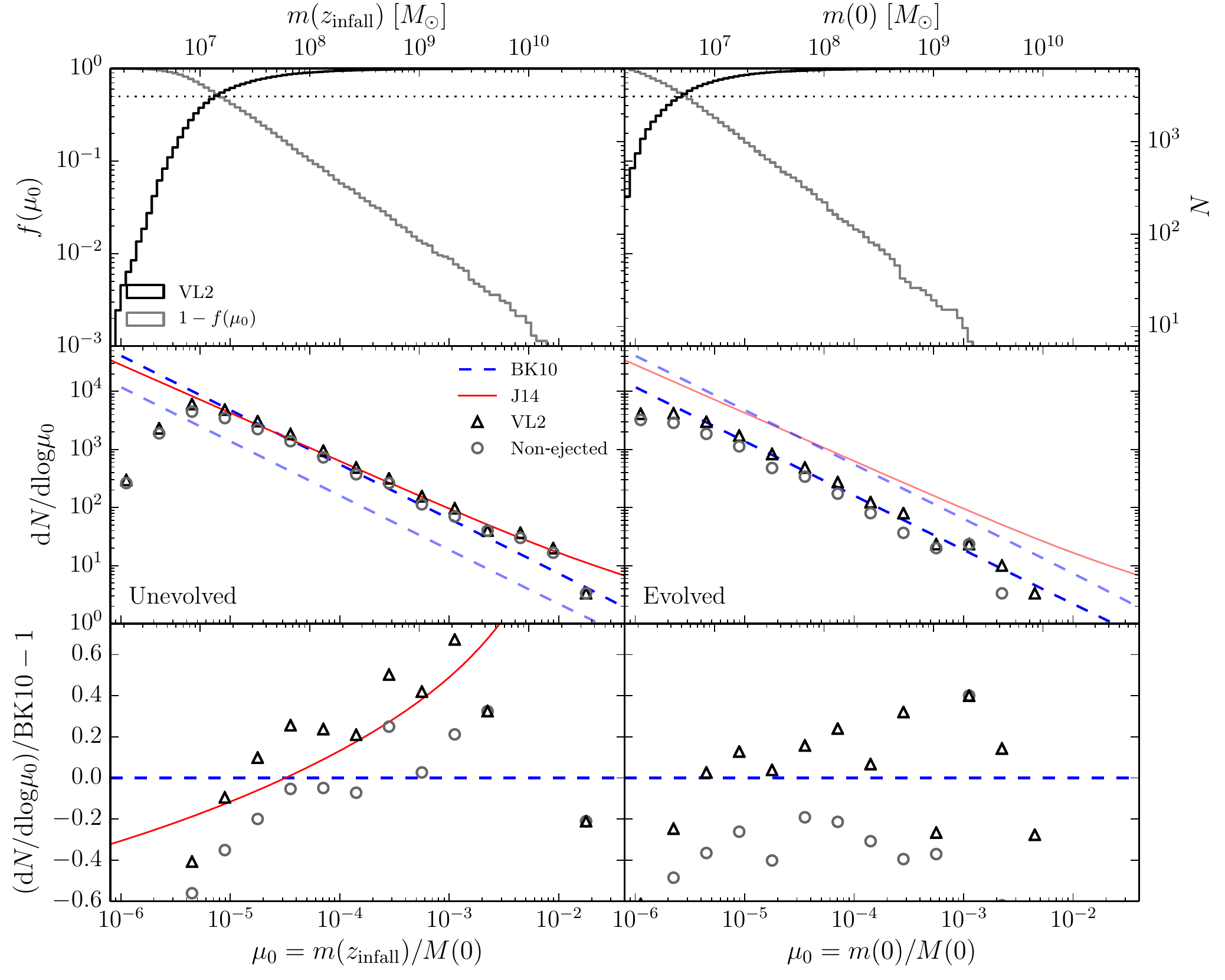}
\end{center}
\caption{Top panels show the cumulative distribution in $\mu_0$ measured at infall (left) and at $z = 0$ (right) for all 6145 VL2 subhalos. The gray histogram shows the reverse cumulative distribution function. Middle panels show the corresponding unevolved (left) and evolved (right) mass functions for all subhalos (black triangles) and only ``non-ejected'' subhalos (gray circles). Hence, the gray circles correspond to only those 4607 subhalos currently residing within $R_{\rm vir}$ at $z = 0$ (see \S\ref{subsec:subdef}). This is done for the purpose of comparing to the Aquarius simulation, shown as the dark dashed blue line, based on the BK10 fitting function. The VL2 and Aquarius mass functions show only those subhalos that survive to the present epoch. In contrast, the solid red line traces the fitting function of \citet{jiang/vdbA:2014} for the unevolved mass function of {\em all} subhalos accreting onto the host. The lightly shaded curves in the left (right) panel correspond to evolved (unevolved) quantities in order to better show the difference between the two mass functions. The bottom left (right) panel shows the relative difference between the various data and the unevolved (evolved) BK10 fitting function.}
\label{fig:dndm} \vspace{0.2cm}
\end{figure*} 

\subsection{Definition of an orbit}
\label{subsec:orbdef}

In \S\ref{sec:results4} we compute subhalo quantities, such as tidal mass, taken over the course of an orbital period. To do so requires a precise definition of an ``orbit". This is a complicated task since an orbit within a spherical potential is not closed, generally, and traces a rosette pattern, oscillating radially between a minimum pericenter, $\rperi$, and maximum apocenter, $\rapo$ \citep[see, e.g.,][]{binney/tremaine:1987}. Furthermore, the mass distribution in realistic halos departs significantly from spherical symmetry, due both to triaxiality in the smooth component, as well as substructure. Finally, subhalos slowly spiral inward due to dynamical friction both from the background matter distribution as well as stripped material. Consequently, energy and angular momentum are not in general conserved, and we require a robust and physical definition of an orbit that does not depend on simplifying assumptions such as spherical symmetry and conserved quantities.

We choose to work solely from knowledge of the radial position of the subhalo as a function of time, determining the local minima (pericenters) and maxima (apocenters). Due to the somewhat coarse time information, apocenters are generally more accurately determined than pericenters, since halos spend a larger fraction of time further away from the host center. Thus, we define an orbit as that segment of the subhalo trajectory between two successive apocenters. A given orbit is therefore characterized by the time at first and last apocenters $t_1$ and $t_2$, the two apocenters $r_{\rm apo,1}$ and $r_{\rm apo,2}$, and the pericenter, $r_{\rm peri}$. We take the mean of the two apocenters, $\overline{r}_{\rm apo}\equiv (r_{\rm apo,1}+r_{\rm apo,2})/2$, and define an effective eccentricity
\bq
e_{\rm eff} \equiv \frac{\overline{r}_{\rm apo}-r_{\rm peri}}{\overline{r}_{\rm apo}+r_{\rm peri}},
\label{eq:eff}
\eq
while the period of the orbit is $t_{\rm orb} = t_2 - t_1$. 


\section{Results}
\label{sec:results}

\subsection{Host halo}
\label{sec:results1}

\begin{figure} \smjustify
\includegraphics[width=\smhwidth]{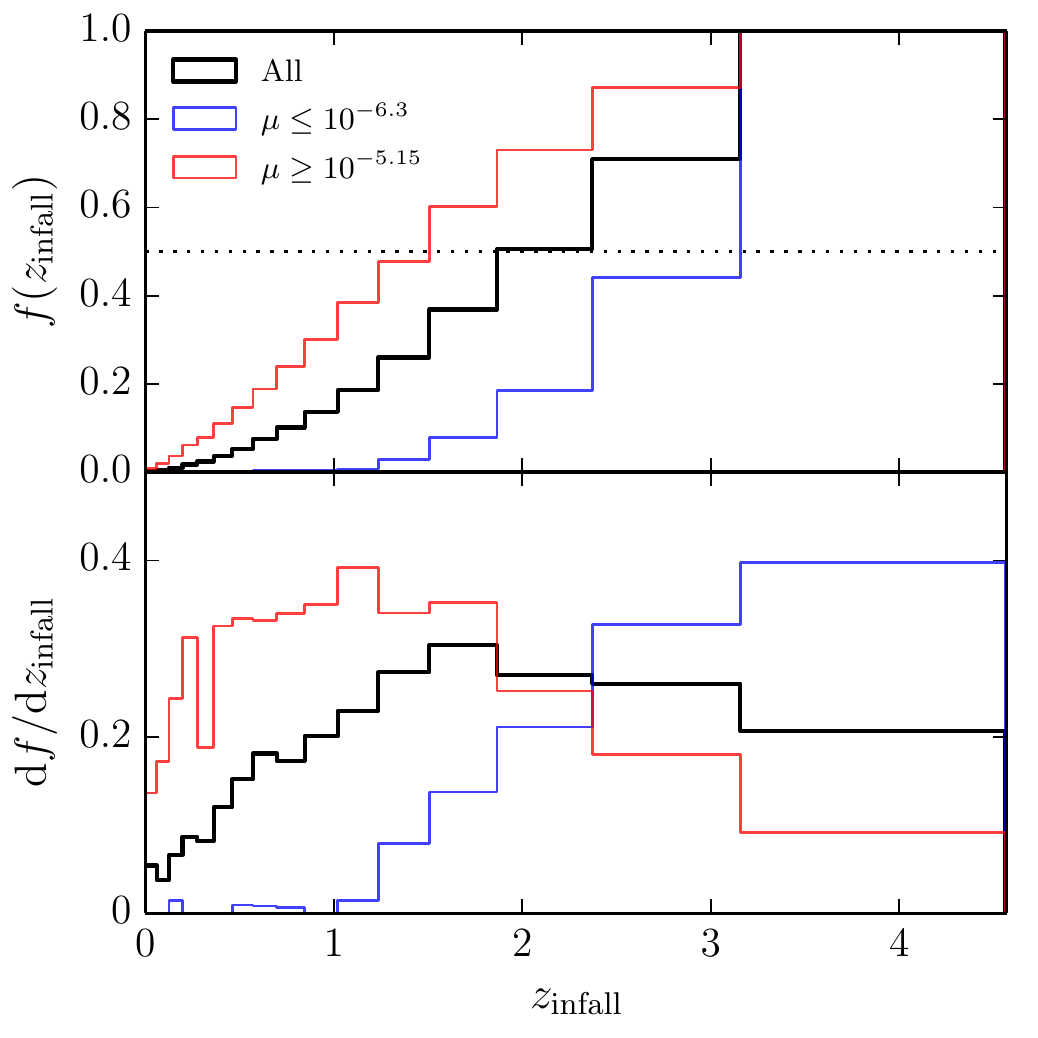}
\caption{Cumulative (top panel) and differential (bottom panel) infall redshift distributions for bins equally spaced in cosmic time. The black histogram traces the total sample of subhalos while the blue and red histograms show distributions for the $1\sigma$ outliers with the smallest and largest present-day mass ratios, respectively.}
\label{fig:zinfall} \vspace{0.2cm}
\end{figure} 

We begin by presenting the derived properties of the host halo using the method outlined in \S\ref{subsec:vl2con}. Figure \ref{fig:mwh} shows the redshift evolution of the host virial mass and concentration. Open circles denote the redshifts for which the VL2 catalogues are sampled while the solid black lines trace the result we derive after performing a cubic spline interpolation on the time evolution of $R_{\rm max}$ and $V_{\rm max}$. Our method finds the host to evolve from a virial mass of $M_{\rm vir} = \unit{1.7 \times 10^{11}}{\mass}$ at $z = 4.56$ to $M_{\rm vir} = \unit{1.9 \times 10^{12}}{\mass}$ at $z = 0$. The concentration evolves from $C_{\rm vir} = 2.6$ at early times to $C_{\rm vir} = 12.2$ at the present day. 

\citet{wechsler/etal:2002} showed that halo concentration is strongly related to mass assembly history. In particular, evolution in concentration and virial mass can be described remarkably well using a single parameter, $a_c = 1/(1+z_c)$, defined as the formation or collapse time of the halo. They provide fitting relations $C_{\rm vir} = 4.1a/a_c$ and $M_{\rm vir}(z) = M_{\rm vir}(0) {\rm exp}[-2a_cz]$ which we plot in Figure \ref{fig:mwh} spanning the redshift range $z_c = 1.7 - 3.7$. The VL2 data fits well within the shaded region which may reflect an initial collapse time of $z \sim 3.7$ followed later by an episode of significant mass assembly at $z \sim 1.7$ which resets the concentration back down to $C_{\rm vir} \sim 4$.

\subsection{Mass functions}
\label{sec:results2}

The subhalo mass function provides a statistical measure of the amount of substructure within a host as a function of mass scale. In general, we can speak of two subhalo mass functions: the {\em unevovled} and {\em evolved} mass functions. The unevolved mass function counts the number of subhalos based on their mass at the time of infall. The choice of name emphasizes that this is a measure of the distribution of subhalos before they have had time to evolve under the influence of tidal processes within the host. The evolved mass function, on the other hand, counts the number of subhalos based on their present-day masses. 

In the middle panels of Figure \ref{fig:dndm} we plot both the unevolved and evolved subhalo mass functions measured from VL2. We compare these to the corresponding mass functions from \citet[][BK10]{boylan-kolchin/etal:2010} which were fitted from the Millennium II \citep{boylan-kolchin/etal:2009} and Aquarius \citep{springel/etal:2008} simulations. This is a useful comparison since BK10 considered host halos with similar masses to the VL2 host halo and use the same definition of $z_{\rm infall}$ as we do here. One difference, however, is that BK10 do not consider subhalos that reside outside the virial radius of the host at $z = 0$ (i.e., ejected subhalos; see \S\ref{subsec:subdef}). For more of a direct comparison, we also plot the VL2 mass functions with ejected subhalos removed. The bottom panels show more closely the comparison between VL2 and BK10. 

The VL2 and BK10 unevolved mass functions agree well with each other over most of the mass range seen here. The sharp cutoff at small mass simply reflects the resolution limit of VL2. There is considerable disagreement at the high mass end, though this regime is inherently noisy due to small number statistics. This can be seen in the top left panel where the cumulative distribution in $\mu_0$ at infall is shown; only 11 objects with $\mu_0 > 4 \times 10^{-3}$ at infall exist. Including ejected subhalos enhances the VL2 unevolved mass function by a constant factor indicating that infall mass does not play a significant role in determining whether a subhalo resides outside of $R_{\rm vir}$ at $z = 0$.

Note that the unevolved mass functions shown here correspond only to those subhalos that accrete onto the host and remain intact at $z = 0$. The red line in Figure \ref{fig:dndm} traces equation (21) of \citet{jiang/vdbA:2014} which shows the unevolved mass function for {\em all} subhalos ever accreted onto the host. This mass function is found to have a universal form \citep{vandenbosch/etal:2005,giocoli/etal:2008,li/mo:2009}, independent of host halo mass and cosmology, except perhaps a small dependence on $n_s$ \citep{yang/etal:2011}. The main difference between this mass function and that of surviving subhalos occurs at high $\mu$ where dynamical friction selectively disrupts massive subhalos after infall. The VL2 unevolved mass function (black triangles) agrees well with the red line albeit with a small systematic shift upwards. In \S\ref{sec:dispop} we analyze disrupted subhalos in VL2 and find that including them here would further boost the black triangles upward by $\sim 5\%$ in the range $10^{-5} \lesssim \mu_0 \lesssim 10^{-3}$ (see Figure \ref{fig:fdisrpt}). We are indeed focusing on the low-mass regime where dynamical friction and tidal disruption are relatively unimportant for the vast majority of subhalos.

The right panels of Figure \ref{fig:dndm} show the evolved counterparts. The evolved mass function can be thought of as a shift to lower mass due to tidal stripping. This can be seen by comparing the dark and lightly shaded blue lines. The VL2 evolved mass function lies systematically below the BK10 result at a level of about $30\%$. BK10 quote an intrinsic halo-to-halo scatter of $18\%$ for $\mu_0 \lesssim 10^{-3}$ which is not enough to explain the discrepancy seen here. Instead, the difference observed here is most likely related to differences in cosmological parameters. In particular, VL2 uses $\sigma_8 = 0.74$ while Aquarius simulates larger perturbations with $\sigma_8 = 0.9$. It is not straightforward to describe how this difference manifests in the evolved mass function since there are at least two competing effects. On the one hand, the lower amplitude of fluctuations in VL2 will yield later formation times meaning that subhalos have less time on average to lose mass since infall. On the other hand, later formation times also yield lower subhalo concentrations which promote more efficient mass loss (see \S\ref{subsec:resmloss}).

\begin{figure}[t] \smjustify
\includegraphics[width=\smhwidth]{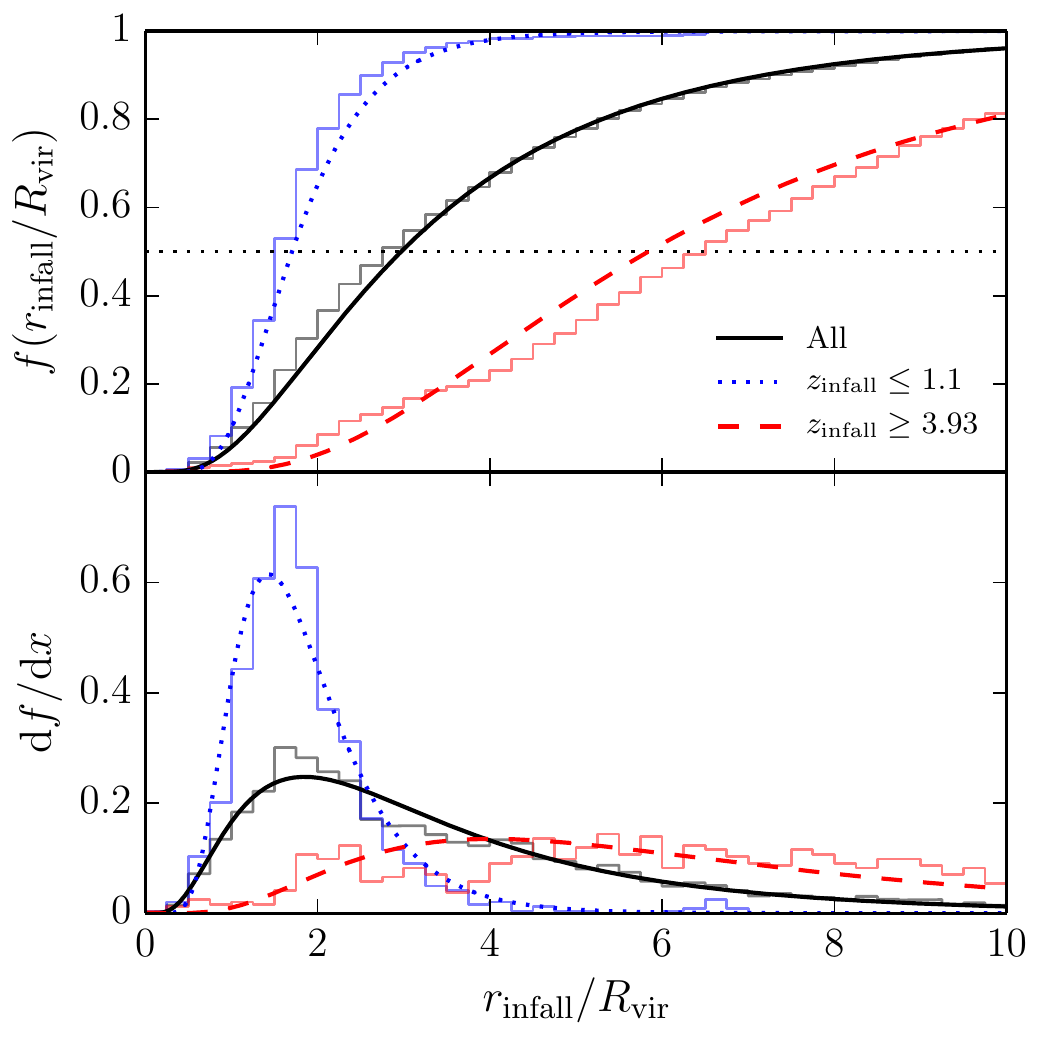}
\caption{Cumulative (top panel) and differential (bottom panel) distributions in the ratio of the radial distance at infall, $r_{\rm infall}$, to the virial radius of the host, $R_{\rm vir}$, at that time. The black histogram traces the total sample of subhalos while the blue and red histograms show distributions for the $1\sigma$ outliers with the smallest and largest infall redshifts, respectively. The solid black curve, dotted blue curve, and dashed red curve trace lognormal fits to the black, blue, and red histograms, respectively. The mean and standard deviation of these fits are given in Table \ref{table:lognorms}.}
\label{fig:rinfall} \vspace{0.2cm}
\end{figure} 

Another factor that may contribute to this difference lies in the definition of tidal mass used by VL2. As described in \S\ref{sec:vl2data}, subhalo masses in VL2 are underestimated at pericenter due to the simplified scheme used in computing mass based on local density comparisons. This is in contrast to the unbinding procedure used by Aquarius with the code {\small SUBFIND} \citep{springel/etal:2001}. As a result, the VL2 evolved mass function will be biased toward smaller masses as some subhalos will be found near pericenter at $z = 0$ (the unevolved mass function is less affected since subhalos are generally near apocenter at infall). Nevertheless, we only expect this to be a partial effect since a suppression of $30\%$ was also seen in \citet{klypin/etal:2011} when comparing VL2 and Aquarius $v_{\rm max}$ functions. The physical mechanism leading to the systematic difference between the VL2 and Aquarius evolved mass functions remains to be seen. 

\subsection{Subhalo properties at infall}
\label{sec:results3}

In this section we focus on subhalo statistics at the time of infall onto the host. In particular, we investigate the redshift at which infall occurs (\S\ref{sec:results-zinf}), the radial distance from the host at which tidal truncation initiates (\S\ref{sec:results-rinf}), and show distributions in orbital energy (\S\ref{sec:results-etainf}) and angular momentum (\S\ref{sec:results-epsinf}) at infall. The results presented here are important as inputs into semi-analytic models of substructure evolution and extend the results of previous works to much lower mass.

\subsubsection{Redshift: $z_{\rm infall}$}
\label{sec:results-zinf}

In Figure \ref{fig:zinfall} we show the cumulative distribution of infall redshift for all 6145 subhalos with $z_{\rm infall} \leq 4.56$. We see that half of the population has fallen into the host by $z = 2$. We also plot separate distributions for the $1\sigma$ outliers having the smallest $16\%$ present-day mass ($\mu \leq 5 \times 10^{-7}$) and largest $16\%$ present-day mass ($\mu \geq 7 \times 10^{-6}$). We see that presently more massive subhalos tend to have fallen in at more recent times. There are two reasons for this trend: (i) structure forms hierarchically, so halos falling in at earlier times were on average less massive to begin with than those infalling later; (ii) subhalos of a given mass that fell in earlier have had more time to undergo tidal stripping, and will be less massive today.

\subsubsection{Radius: $r_{\rm infall}$}
\label{sec:results-rinf}

\begin{figure} 
\begin{center}
\includegraphics[width=\smhwidth]{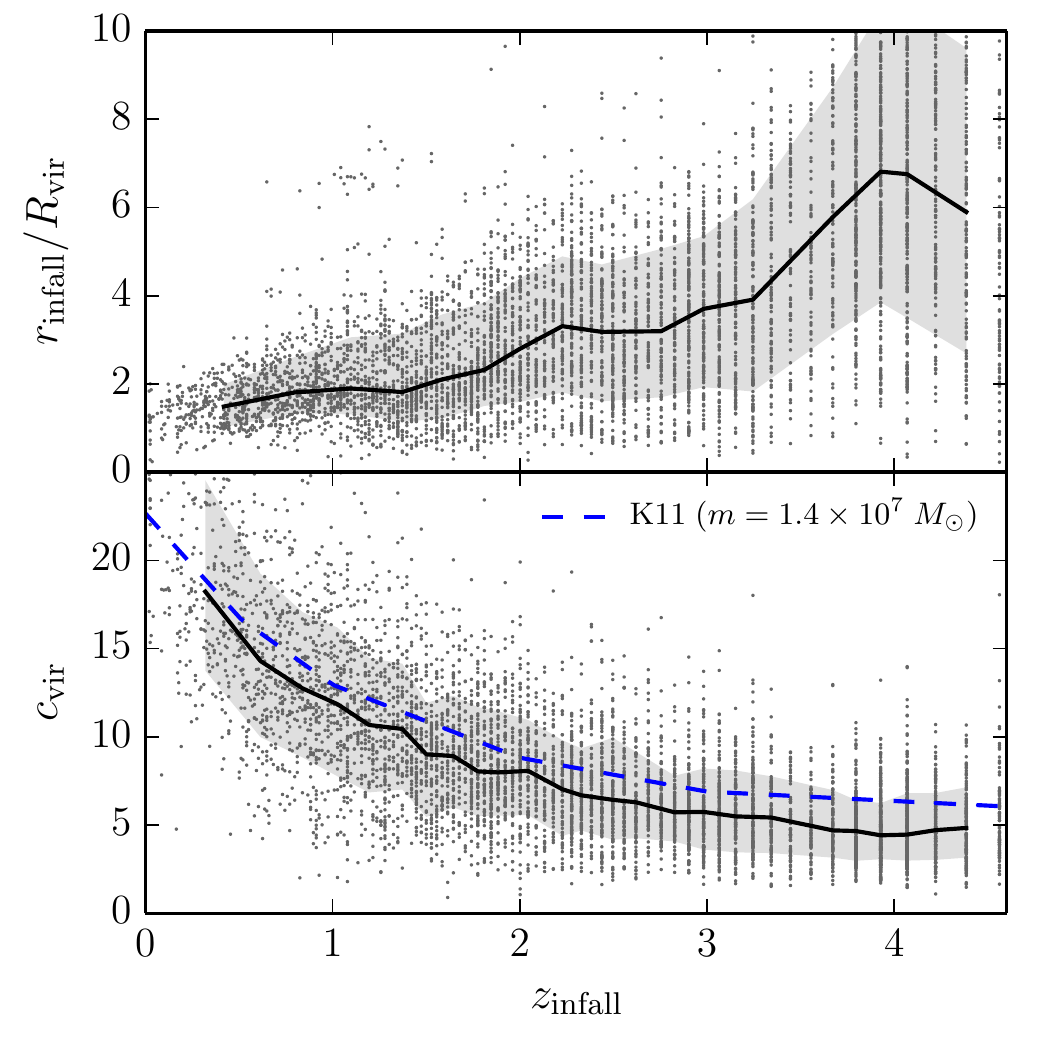}
\vspace{-0.8cm}
\end{center}
\caption{Top panel shows the radius--redshift relationship at infall, for each subhalo (points), with median (solid line) and $1\sigma$ distribution (shaded region) in each $z_{\rm infall}$ bin. Bottom panel shows the concentration--redshift relationship at infall. Solid line and and shaded region have the same meanings as in the top panel. Shown also is the mean concentration--redshift relationship from Klypin et al. (2011) for a subhalo mass of $1.4 \times 10^7 M_\odot$, which we find to be the median infall mass, independent of redshift. Halos falling in earlier are less concentrated {\em and} have their growth truncated at larger distances.}
\label{fig:crvz} \vspace{0.2cm}
\end{figure}

Figure \ref{fig:rinfall} shows the distribution of the radial distance, $r_{\rm infall}$, between the subhalo and host at infall normalized to the virial radius of the host at that time. Hence, we are plotting the relative distance at which the subhalo has its growth history truncated due to tidal interactions with the host. We also plot separate distributions for the $1\sigma$ outliers with the most recent infall, $z_{\rm infall} \leq 1.1$, and earliest infall, $z_{\rm infall} \geq 3.93$. In each case, the differential distribution can be well approximated by a lognormal form in $r_{\rm infall}/R_{\rm vir}$. The mean and standard deviation of the least-squared lognormal distribution for each population are summarized in Table \ref{table:lognorms}. 

Somewhat surprisingly, we find that over 90 per cent of subhalos undergo tidal growth truncation outside of the virial radius, with roughly 50 per cent infalling at a distance of more than three virial radii from the host.  Considering halos falling in at the earliest times, $z_{\rm infall} > 3.93$, this fraction rises above 80 per cent. We emphasize, however, that these numbers are likely biased toward larger values since our definition of infall does not exclude the possibility of group preprocessing for which tidal truncation first occurs via interactions with halos other than the future host. Nevertheless, our findings are in qualitative agreement with past studies \citep{hahn/etal:2009,behroozi/etal:2014} showing that tidal truncation generally occurs outside of $R_{\rm vir}$. This trend is also apparent in the top panel of Figure \ref{fig:crvz}, where we show the relationship between infall radius and redshift directly. {\em Why do halos at high redshift begin to be affected so far outside the host?}

\begin{table}[b] 
\caption{Lognormal $r_{\rm infall}/R_{\rm vir}$ Fits.}
\centering
\begin{tabular}{c c c}
\hline\hline
Population & $\mu$ & $\sigma$ \\
\hline
All subhalos & 1.09 & 0.69 \\
$z_{\rm infall} \leq 1.1$ & 0.53 & 0.42 \\
$z_{\rm infall} \geq 3.93$ & 1.76 & 0.61 \\
\label{table:lognorms}
\end{tabular}
\end{table}

A first hint is provided upon inspection of the bottom panel of Figure \ref{fig:crvz}, where the infall concentration is plotted against infall redshift. We see a strong correlation, with halos that fall in earlier having much lower concentrations. This is expected because the typical mass of an infalling halo does not change very strongly with redshift. Thus, the concentrations of infalling halos grow roughly as expected for halos of a fixed mass, e.g., $1.4 \times 10^7 M_\odot$, which we find to be the median infall mass, independent of $z_{\rm infall}$ (see also Figures \ref{fig:dndm} and \ref{fig:fdisrpt}). This is shown as the blue dashed line, which is the mean concentration--redshift relationship at fixed mass as determined by \citet{klypin/etal:2011}\footnote{Note that the concentration-redshift relation observed here is systematically lower than the \citet{klypin/etal:2011} curve. One reason is that the latter was calibrated against the Bolshoi simulation which had an enhanced amplitude of perturbations with $\sigma_8 = 0.82$ compared to the value of $0.74$ employed by VL2. At a fixed redshift we expect VL2 concentrations to be lower than in the Bolshoi simulation based on the notion that concentration reflects the background density of the Universe at the time of halo formation.}. It would seem that a plausible explanation lies in the concentration of infalling halos.

\begin{figure} 
\begin{center}
\includegraphics[width=\smhwidth]{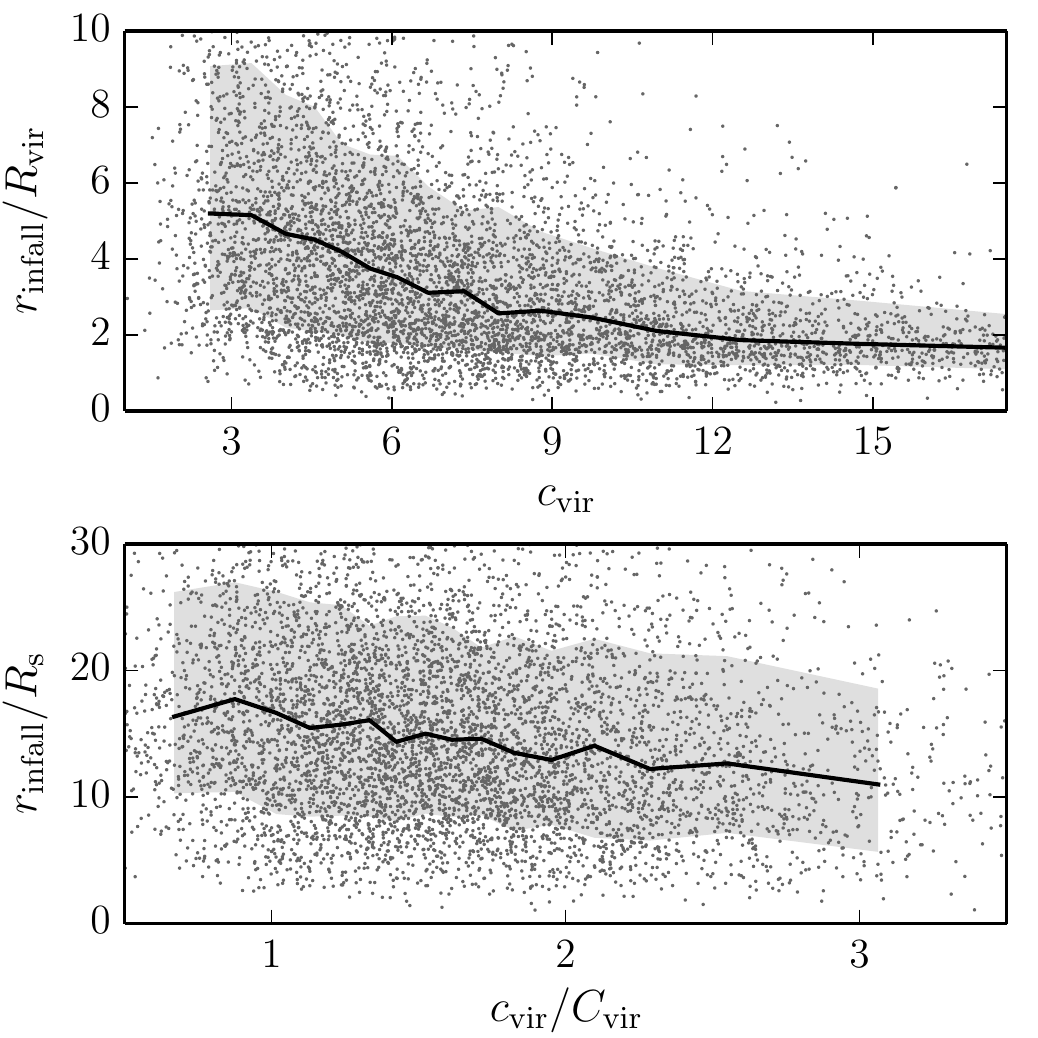}
\vspace{-0.8cm}
\end{center}
\caption{Top panel shows the infall radius--concentration relationship at infall. The bottom panel shows a similar result with $c_{\rm vir}$ and $r_{\rm infall}$ normalized to the concentration and scale radius of the host, respectively. In both panels the solid line and shaded region have the same meaning as in Figure \ref{fig:crvz}.}
\label{fig:cvr} \vspace{0.2cm}
\end{figure}

This can be directly tested by plotting $r_{\rm infall}/R_{\rm vir}$ versus $c_{\rm vir}$, as shown in the top panel of Figure \ref{fig:cvr}. We see a definite trend of more concentrated subhalos coming closer to the host before undergoing infall. Our physical interpretation is that highly concentrated subhalos with compact density profiles are more resilient to tidal stripping. Hence, they are able to plunge deeper into the potential well of the host before appreciable mass loss occurs. Note that the relationship seen in Figure \ref{fig:cvr} appears weaker than the trend observed when comparing $c_{\rm vir}$ versus $z_{\rm infall}$ in Figure \ref{fig:crvz}. In particular, there is a significant fraction of halos with $c_{\rm vir}<10$ and $r_{\rm infall}/R_{\rm vir}<3$, with the vast majority of these falling in at late times. This implies there are other effects at high redshift that hinder the growth of infalling halos, in addition to lower central densities. 

Before advancing we note that a trend in $r_{\rm infall}/R_{\rm vir}$ versus $c_{\rm vir}$ is expected to exist even if $r_{\rm infall}$ does not change much with time. This is based on the fact that both $c_{\rm vir}$ and $R_{\rm vir}$ generally increase with time due to the expansion of the universe. To try to account for this, we plot, in the bottom panel of Figure \ref{fig:cvr},  $r_{\rm infall}/R_s$ versus $c_{\rm vir}/C_{\rm vir}$, where $R_s = R_{\rm vir}/C_{\rm vir}$ is the scale radius of the host. Normalizing $c_{\rm vir}$ and $r_{\rm infall}$ in this way acts to remove redshift dependencies in $c_{\rm vir}$ and $R_{\rm vir}$. We are thus plotting how close tidal truncation occurs relative to the central density peak of the host as a function of subhalo concentration relative to that of the host. Although weaker, we still see a definite trend of more concentrated subhalos coming closer to the central depth of the host before experiencing tidal truncation. This lends support to the notion of an intrinsic radius-concentration relation for infalling subhalos.

\subsubsection{Orbital energy: $\eta_{\rm infall}$}
\label{sec:results-etainf}

\begin{figure} \smjustify
\includegraphics[width=\smhwidth]{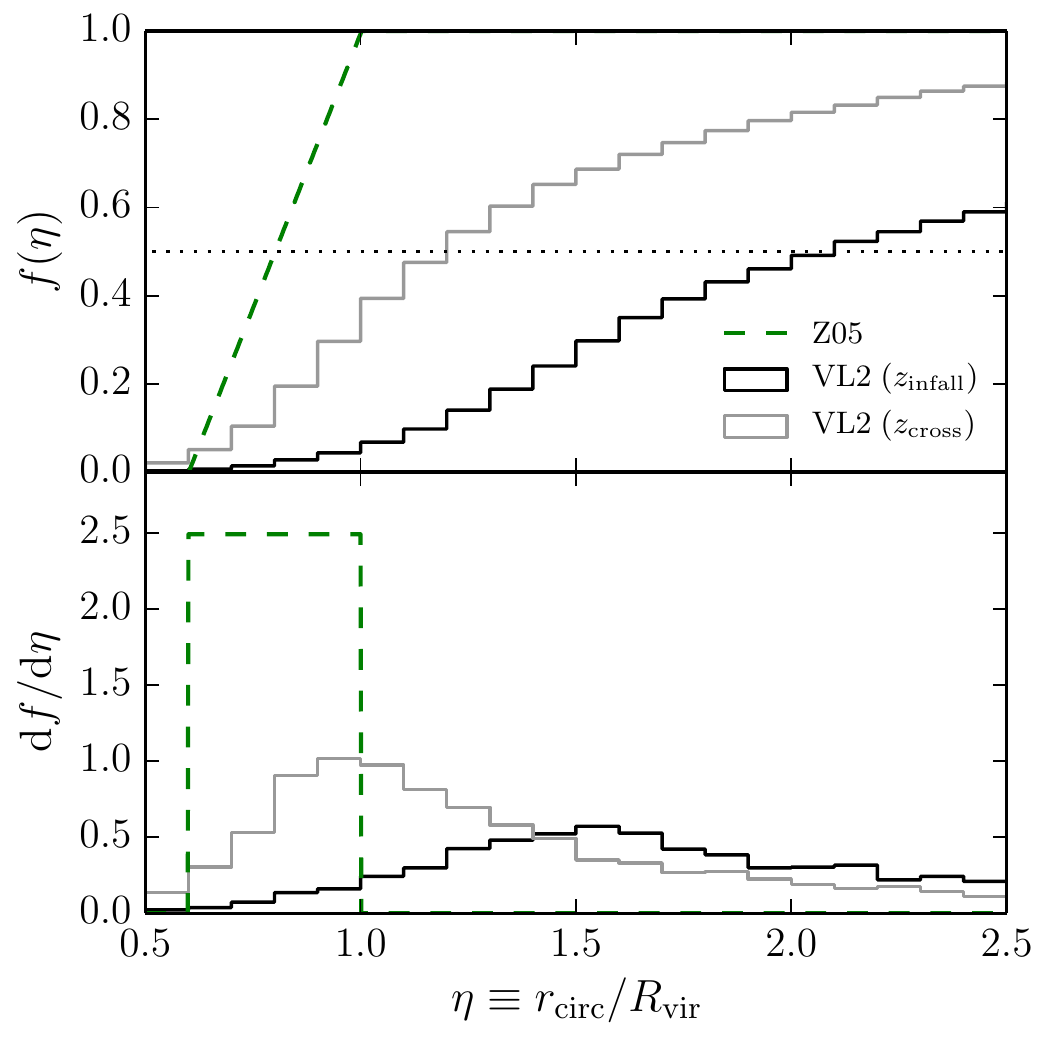}
\caption{Cumulative (top panel) and differential (bottom panel) VL2 distributions in $\eta$ at the time of infall (black histogram) and first virial crossing with a point mass potential (gray histogram). Compared to the latter is the green dashed curve showing the uniform distribution reported by Z05.}
\label{fig:eta} \vspace{0.2cm}
\end{figure} 

Semi-analytic models of substructure evolution require two inputs as initial conditions for subhalo orbits: energy and angular momentum. In this section we present the infall distribution of energy as seen in VL2 and proceed in the next section with angular momentum. In accordance with past studies, we parametrize the infall energy in terms of the variable $\eta \equiv r_{\rm circ}/R_{\rm vir}$, where $r_{\rm circ}$ is the radius of the circular orbit of the same energy as the subhalo and $R_{\rm vir}$ is the virial radius of the host at infall. We compute this quantitiy by first evaluating equation (\ref{eq:orben}) for the orbital energy, and then solving equation (\ref{eq:orbrcirc}) for $r_{\rm circ}$ based on an orbit within an isolated NFW host potential.

This parametrization is only valid for subhalos on bound orbits ($E < 0$). It turns out that this condition is not very restrictive since only 38 ($0.6\%$) of the 6145 subhalos are on unbound orbits at the time of infall. This number is still small at $z = 0$ when only 75 subhalos are found to be on unbound orbits. Interestingly, {\em all} 75 of these subhalos are outside of $R_{\rm vir}$ at $z = 0$ (i.e., they are ejected subhalos) meaning that {\em no} subhalos within the present virial radius are unbound. In contrast, only a small fraction (7/38) of unbound subhalos at infall end up being part of the ejected population of subhalos at $z = 0$. Moreover, {\em none} of the unbound subhalos at infall are also unbound at $z = 0$. Being unbound from the host potential at infall correlates neither with being presently unbound nor with being found outside $R_{\rm vir}$ at $z = 0$. Instead, it is likely that presently unbound subhalos acquire orbital energy from gravitation interactions after infall. This can be achieved, for example, through three-body interactions between merging groups of subhalos as they make their first passage together around the host \citep{sales/etal:2007,ludlow/etal:2009}.

In Figure \ref{fig:eta} we plot the VL2 distribution in $\eta$ at infall for all bound subhalos. This is compared to the uniform distribution between [0.6, 1] used in the semi-analytic model of \citet[][Z05]{zentner/etal:2005} based on the analysis of the N-body simulations of \citet{klypin/etal:2001} and \citet{kravtsov/etal:2004}. This also serves as the basis for the input distributions of $\eta$ used in the semi-analytic models of \citet{gan/etal:2010} and \citet{jiang/vandenbosch:2014}. The VL2 result, with a peak at $\eta \sim 1.5$, is in clear disagreement with the Z05 distribution. However, as described in \S\ref{subsec:eldef}, our calculation of $\eta$ is not directly comparable to that of Z05. Firstly, Z05 report $\eta$ at the time a subhalo first crosses through $R_{\rm vir}$, which occurs at much later times on average than $z_{\rm infall}$ (see \S\ref{sec:results-rinf}). In addition, orbital energy and $r_{\rm circ}$ are computed via equations (\ref{eq:orbeng-sa}) and (\ref{eq:orbcirc-sa}), valid for a point mass host potential. The gray histogram in Figure \ref{fig:eta} shows the result of applying this method to the VL2 data. The agreement with Z05 is better, but still heavily offset toward larger values of $\eta$.

\begin{table}[t] 
\caption{Lognormal ${\rm d}f/{\rm d}\eta$ Fits.}
\centering
\begin{tabular}{c c c c c}
\hline\hline
 & \multicolumn{2}{c}{$z_{\rm infall}$ with NFW host} & \multicolumn{2}{c}{$z_{\rm cross}$ with point host} \\
Population & $\mu$ & $\sigma$ & $\mu$ & $\sigma$ \\
\hline
All subhalos & 0.70 & 0.50 & 0.18 & 0.40 \\
$ z \leq 1$ & 0.38 & 0.21 & 0.29 & 0.41 \\
$1 \leq z \leq 2$ & 0.53 & 0.36 & 0.22 & 0.38 \\
$2 \leq z \leq 3$ & 0.88 & 0.44 & 0.17 & 0.39 \\
$z \geq 3 $ & 1.50 & 0.54 & 0.10 & 0.40 \\
\label{table:lognorms-eta}
\end{tabular}
\end{table}

\begin{figure*}[ht]
\begin{center}
\includegraphics[width=0.95\textwidth]{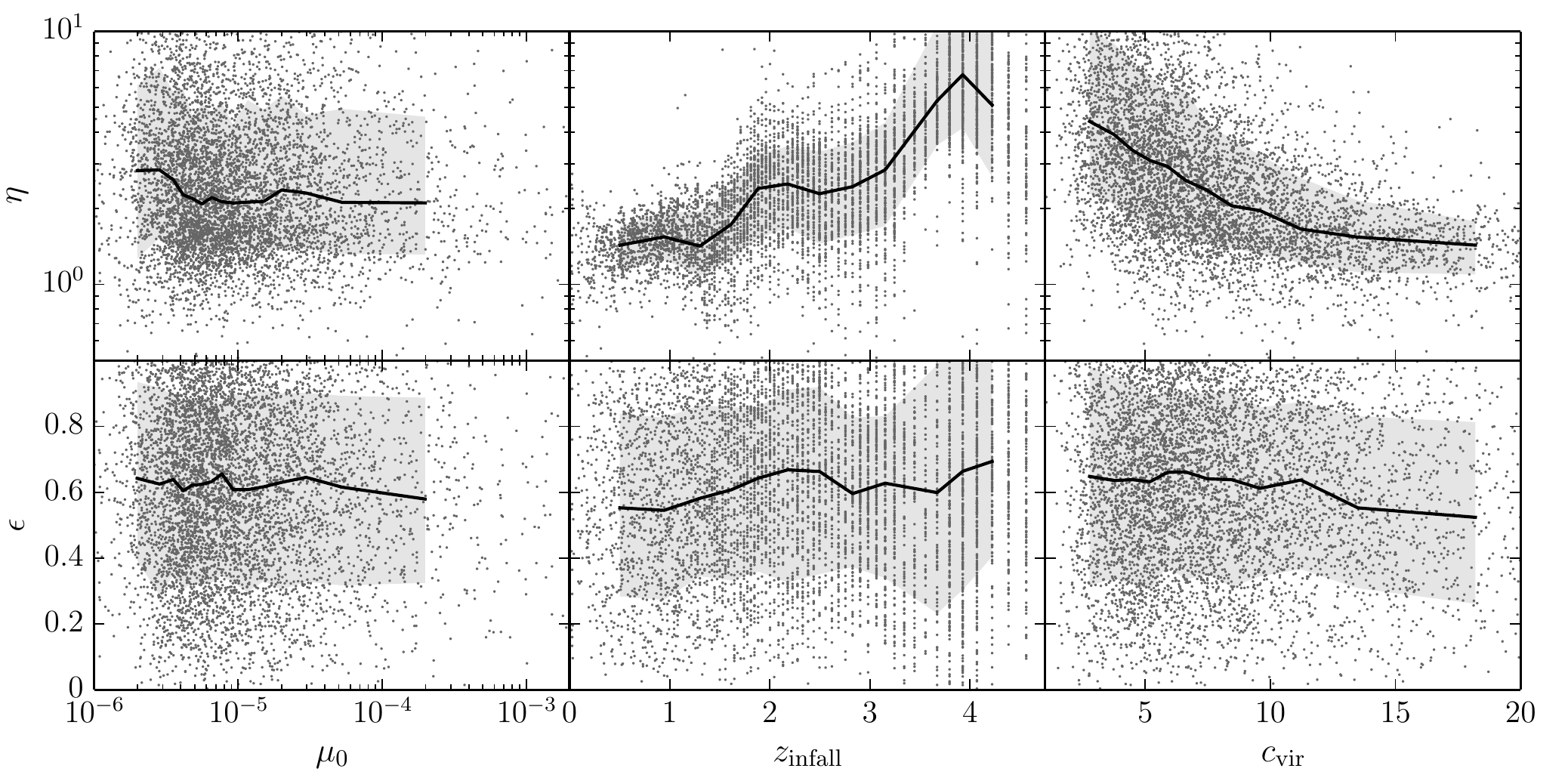}
\end{center}
\caption{The dependence of $\eta$ (top panels) and $\epsilon$ (bottom panels) at infall on mass ratio at infall (left), infall redshift (middle), and concentration at infall (right). In both cases we assume an NFW potential for the host. Points denote individual subhalos while the solid line shows the median trend for bins along the $x$-axis with the corresponding $1\sigma$ spread indicated by the shaded region.}
\label{fig:etaeps} \vspace{0.2cm}
\end{figure*} 

In the recent work of \citet[][J15]{jiang/etal:2014b} it was shown that $\eta$ depends on mass ratio at infall, with smaller objects tending toward larger $\eta$. A direct comparison with their result is difficult, however, due to their choice of a singular isothermal sphere host potential. Nonetheless, their analysis suggests that looking only at subhalos with $\mu_0 \gtrsim 10^{-3}$, similar to those resolved in Z05, would shift the gray histogram in Figure \ref{fig:eta} to the left, presumably in better agreement with the uniform distribution. Since VL2 samples only one host halo, however, we cannot test this explicitly due to insufficient statistics in high-mass subhalos. Nevertheless, the result found here corroborates the work of J15 and indicates that the infall distribution assumed for $\eta$ in various semi-analytic models of substructure evolution may only be strictly valid for relatively massive subhalos. Low-mass subhalos $(\mu_0 \lesssim 10^{-3})$ tend to have more kinetic energy, making them less bound to the host, with lower specific binding energy.

We plot in Figure \ref{fig:etaeps} the dependence of $\eta$ on mass for the range of mass captured in VL2. In this low-mass regime, there does not appear to be any significant trend of $\eta$ with infall mass. Instead, we find a strong trend with infall redshift. This trend is attributed to the fact that $R_{\rm vir}$ is an increasing function of time so that $\eta$ is pushed to larger values at earlier times. This trend is slightly suppressed by a competing evolution in $r_{\rm circ}$ with redshift: we find typical values of $r_{\rm circ}$ increase by a factor of $\sim 2$ from $z = 4$ to $z = 0$ (whereas $R_{\rm vir}$ increases by $\sim 8$ in this range). This intrinsic evolution in $r_{\rm circ}$ with $z_{\rm infall}$ indicates that subhalos falling in at earlier times do so on slightly more bound (smaller $r_{\rm circ}$) orbits. The top right panel of Figure \ref{fig:etaeps} shows a strong trend in $\eta$ with $c_{\rm vir}$ which can be attributed to the concentration-redshift relation seen earlier in Figure \ref{fig:crvz}.

We note that the trends observed in the top panels of Figure \ref{fig:etaeps} depend on which version of $\eta$ we present. Namely, we find that using $\eta$ computed at virial crossing for a point mass host potential mostly washes out the dependence of $\eta$ with redshift and concentration. The reason is that subhalos tend to become more bound in time after infall so that $\eta$ is pushed to smaller values at the time of virial crossing, $z_{\rm cross}$, which generally occurs after $z_{\rm infall}$ (see Figure \ref{fig:crvz}). This seems to occur in such a way as to mostly cancel the evolution in $R_{\rm vir}$ with $z$. The difference between the black and gray histograms in Figure \ref{fig:eta} can therefore be mostly explained by two effects: (1) larger $R_{\rm vir}$ at $z_{\rm cross}$ and (2) smaller $r_{\rm circ}$ at $z_{\rm cross}$.

We find that the probability distributions, ${\rm d}f/{\rm d}\eta$, can be well modelled by lognormal distributions. In Table \ref{table:lognorms-eta} we list the best-fit mean and standard deviations for both the total distribution as well as those obtained from different redshift ranges. The latter corresponds to cuts in $z_{\rm infall}$ and $z_{\rm cross}$ for the infall and virial crossing methods, respectively. These fitting functions are appropriate for subhalos with mass ratio $\mu_0 \lesssim 10^{-3}$. Higher mass subhalos should shift closer to the Z05 curve in Figure \ref{fig:eta}. In \S\ref{sec:dispop} we find that subhalos on tightly bound orbits with $\eta < 1$ are preferentially disrupted after infall. However, this has only a small impact on the distributions presented here since this bias is small and there are far fewer disrupted than surviving subhalos. Hence, the fitting functions provided here should be applicable to the total ensemble of subhalos (surviving plus disrupted) that ever accreted onto the host.

\subsubsection{Orbital angular momentum: $\epsilon_{\rm infall}$}
\label{sec:results-epsinf}

Studies of substructure evolution \citep[e.g.,][]{penarrubia/etal:2008} show that subhalos on more radial orbits with lower specific angular momentum plunge deeper into their hosts and experience accelerated mass loss over subhalos on more circular orbits with higher specific angular momentum (see also \S\ref{subsec:resmloss}). Accurately modelling subhalo evolution therefore requires a good handle on the distribution of angular momentum at the time of infall. As such, a great deal of work has been done on measuring this distribution from N-body simulations \citep{navarro/etal:1995,tormen/etal:1997,ghigna/etal:1998,benson:2005,zentner/etal:2005,wang/etal:2005,khochfar/burkert:2006,jiang/etal:2008,wetzel:2011}. The conclusions of these works agree well with each other: the circularity distribution of infalling satellites is peaked at $\bar{\epsilon} \approx 0.5$ and falls off on either side so that neither largely radial ($\epsilon \sim 0$) nor largely circular ($\epsilon \sim 1$) orbits occur. Below we turn our attention to the circularity distribution measured in VL2.

\begin{figure} \smjustify
\includegraphics[width=\smhwidth]{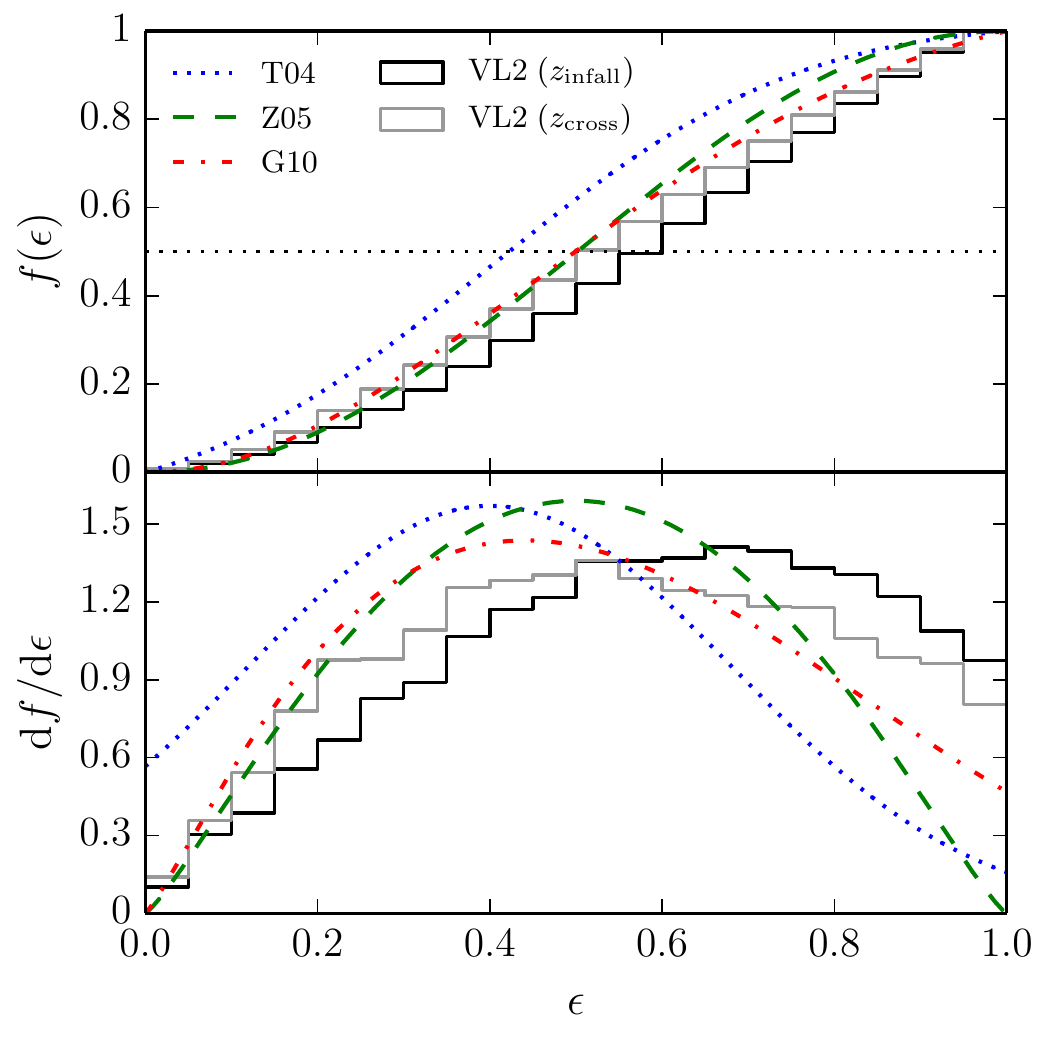}
\caption{Cumulative (top panel) and differential (bottom panel) distributions in $\epsilon$ at the time of infall (black histogram) and first virial crossing with a point mass potential (gray histogram). Compared to the latter are various curves showing the circularity distributions employed in  semi-analytic models of substructure evolution: the blue dotted curve is the distribution from \citet{taylor/babul:2004a}; the green dashed curve is the distribution given by Z05; the red dot-dashed curve is the distribution from \citet{gan/etal:2010}.}
\label{fig:circ} \vspace{0.2cm}
\end{figure}

In Figure \ref{fig:circ} we plot the infall distribution in $\epsilon$ for all bound subhalos. As in Figure \ref{fig:eta}, we show the result at $z_{\rm infall}$ for an isolated NFW host potential (black histogram) as well as the result at first $R_{\rm vir}$ crossing for a point mass host potential. The latter can be compared to the various curves showing the infall distributions used in semi-analytic models of substructure evolution. First, the blue dotted line is the Gaussian distribution used by \citet{taylor/babul:2004a} with mean $\bar{\epsilon} = 0.4$ and standard deviation $\sigma = 0.28$ which was selected so that the final distribution at $z = 0$ matches the results of \citet{tormen/etal:1997} and \citet{ghigna/etal:1998}. Second, the green dashed line shows the one-parameter $\beta$ distribution used in the models of Z05 and \citet{jiang/vandenbosch:2014}. Finally, the red dot-dashed curve shows the infall distribution assumed in the semi-analytic model of \citet{gan/etal:2010}, which was taken from the analysis of the hydrodynamic simulations of \citet{jiang/etal:2008}.

The gray histogram, with median $\epsilon = 0.55$, is in reasonable agreement with the various curves used in semi-analytic models. There is, however, a clear excess in nearly circular orbits with $\epsilon \sim 1$. The reason for this is the same as was discussed in the previous section. It was shown in J15 that circularity is highly dependent on mass, with high mass subhalos tending to move along radial orbits while low mass subhalos tend to have more circular orbits. One possible explanation is related to the environment in which these objects form. High mass halos are more biased towards forming in high density regions such as filaments and are consequently more likely to fall radially into their host with lower specific angular momentum. Low mass subhalos are less biased to forming within filaments and are thus more likely to fall into their host with a larger component of tangential motion. Another possible explanation, independent of the first, is that low-mass subhalos are simply more likely to acquire tangential motion from gravitational interactions with nearby massive objects prior to infall.

In the bottom panels of Figure \ref{fig:etaeps} we show the dependence of $\epsilon$ on mass, redshift, and concentration. As with $\eta$, we find that $\epsilon$ does not exhibit strong mass dependence in the range $\mu_0 \lesssim 10^{-3}$ probed by VL2. There also does not appear to be much dependence on $z_{\rm infall}$, consistent with the earlier work of \citet{wetzel:2011}. There is correspondingly little dependence on $\epsilon$ with infall concentration. Replacing the lower panels of Figure \ref{fig:etaeps} with $\epsilon$ computed at virial crossing for a point mass host potential produces the same (lack of) trends.

Even though the two methods produce similar trends in Figure \ref{fig:etaeps}, it remains difficult to qualitatively describe the differences between the black and gray histograms in Figure \ref{fig:circ}. The reason is the dependence of $L_{\rm circ}$ on the form of the host potential assumed. In fact, we find that $\epsilon$, unlike $\eta$, is very sensitive to the form of the host potential. For example, computing $\epsilon$ at infall with a point mass host potential pushes the black histogram in Figure \ref{fig:circ} to even larger values. Counter intuitively, computing $\epsilon$ at virial crossing with an NFW host potential also pushes the gray histogram to larger values, being almost on top of the black histogram. We therefore note that one should be careful in choosing a fitting function for $\epsilon$ that suits their specific needs.

We fit the circularity distributions in Figure \ref{fig:circ} with the following form:
\bq
\frac{{\rm d}f}{{\rm d}\epsilon} = a \epsilon^\alpha (b - \epsilon)^\beta.
\label{eq:circfitfn}
\eq
We find the best-fit coefficients ($a$, $b$, $\alpha$, $\beta$) = (3.696, 1.12, 1.07, 0.68) at infall with an NFW host potential and ($a$, $b$, $\alpha$, $\beta$) = (1.508, 1.77, 1.05, 2.45) at virial crossing with a point mass host potential. These fitting functions are appropriate for subhalos with mass ratio $\mu_0 \lesssim 10^{-3}$ and are independent of redshift. Higher mass subhalos should have distributions in closer agreement with the other fitting functions plotted in Figure \ref{fig:circ}. In \S\ref{sec:dispop} we show that circularity does not influence the survivability of low-mass subhalos. The fitting functions provided here are thus applicable to the total ensemble of subhalos (surviving plus disrupted) that ever accreted onto the host.

\subsection{Evolution}
\label{sec:results4}

\begin{figure} \smjustify
\includegraphics[width=\smhwidth]{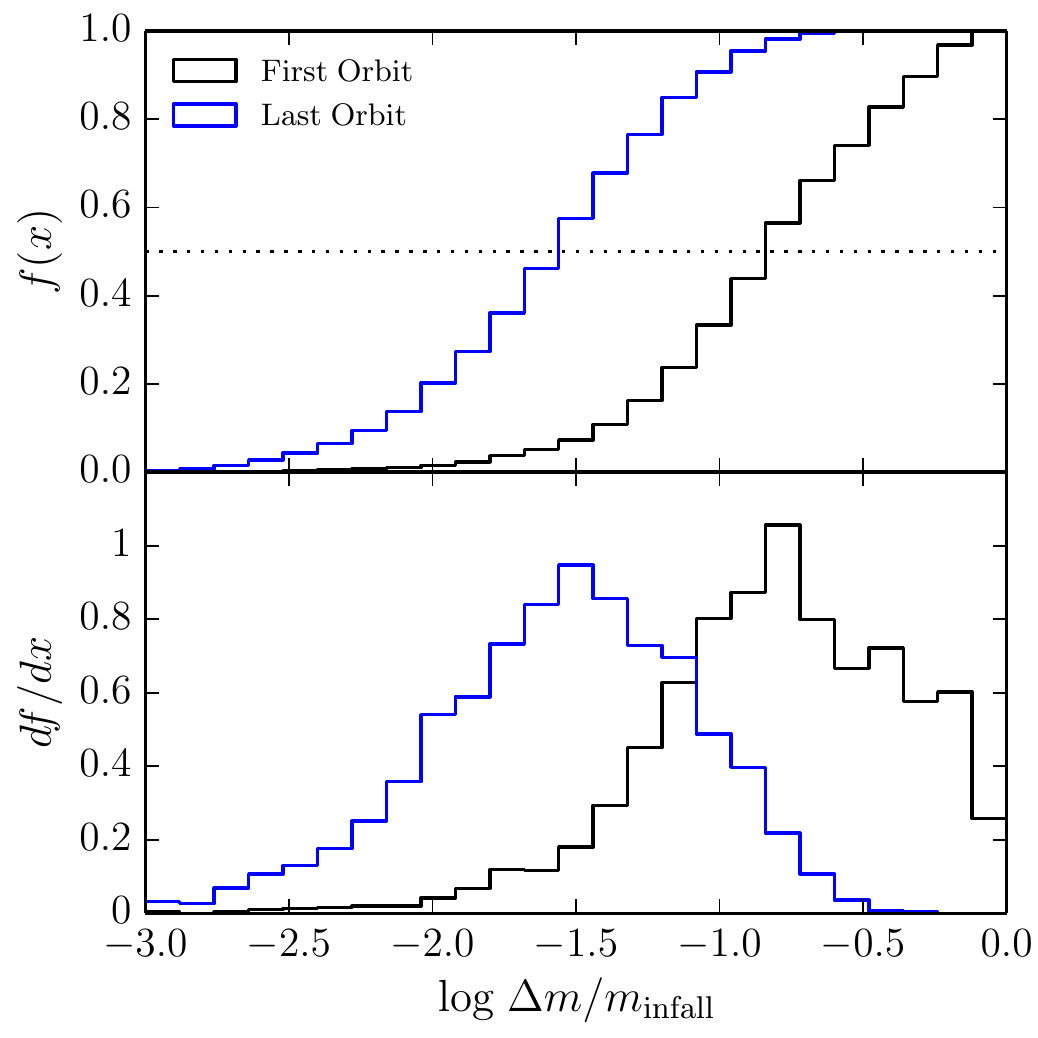}
\caption{Cumulative (top panel) and differential (bottom panel) distributions in the mass loss over the course of an orbit normalized to the mass at infall. The black histogram pertains to the first orbit after infall while the blue histogram shows the last orbit before $z = 0$. Only those 2714 subhalos ($44\%$ of the total population) that complete at least two orbits after infall are shown.}
\label{fig:deltamdist} \vspace{0.2cm}
\end{figure}

In this section we focus on the evolution of subhalo properties over the course of infall to the present day. This includes internal subhalo properties such as tidal mass and central density as well as orbital properties including radial period and angular momentum. Our results are used to test some of the fundamental assumptions underlying models of substructure evolution. 

\begin{figure*} 
\begin{center}
\includegraphics[width=\textwidth]{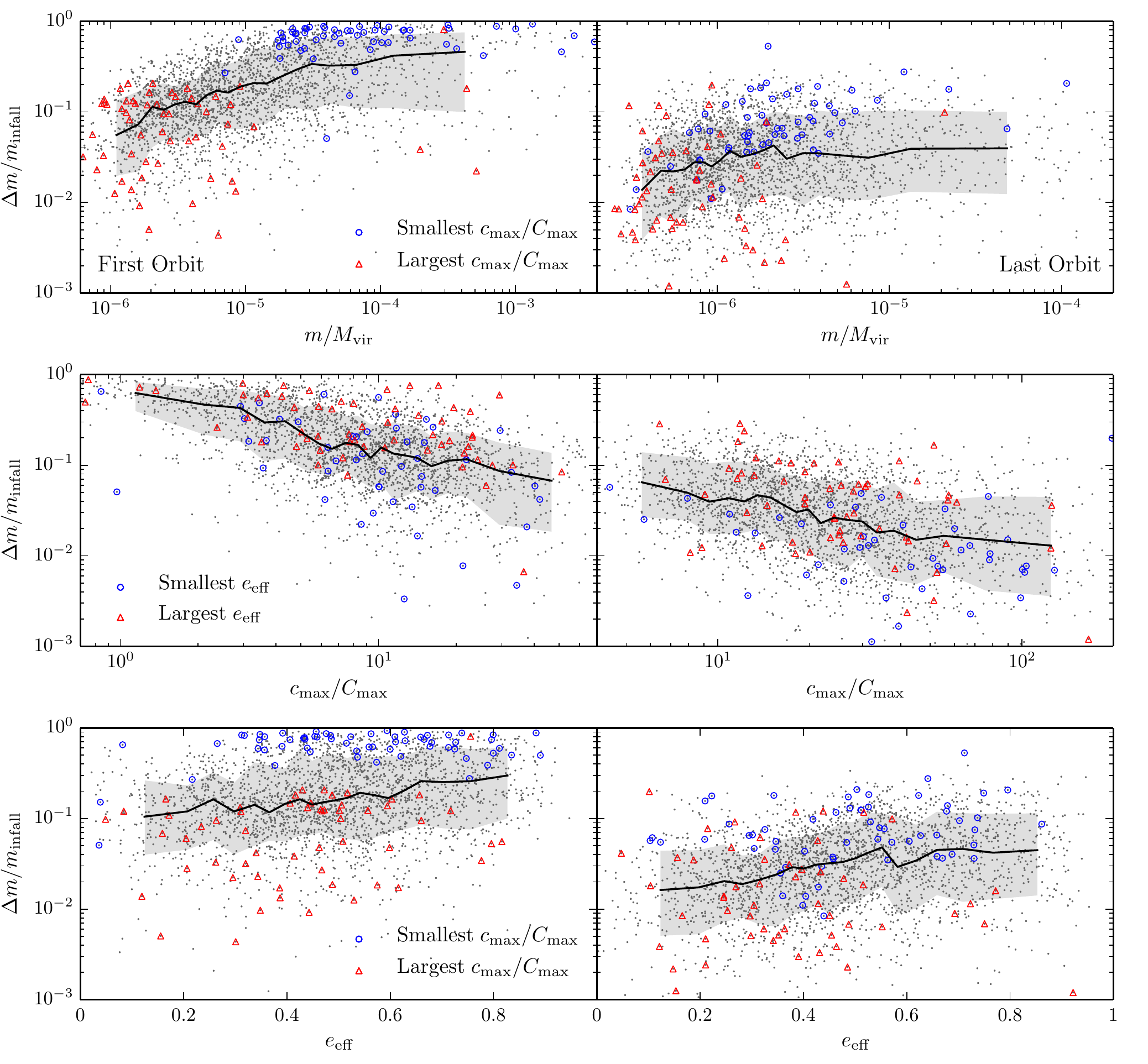}
\end{center}
\caption{Left (right) panels show the fractional amount of mass, $\Delta m$, lost over the course of the first apocenter-to-apocenter orbit after infall (last apocenter-to-apocenter orbit before $z = 0$) normalized to the mass, $m_{\rm infall}$, of the subhalo at the time of infall. The various rows show: (i) fractional mass lost as a function of $m/M_{\rm vir}$ with both quantities computed at the start of the orbit; (ii) fractional mass lost as a function of $c_{\rm max}/C_{\rm max}$ with both quantities computed at the start of the orbit; (iii) fractional mass lost as a function of eccentricity $e_{\rm eff}$ computed from equation (\ref{eq:eff}). In each panel the solid black line traces the median mass loss for bins with equal number of samples in the $x$ axis while the shaded region shows the $1\sigma$ spread about this line. In the top and bottom rows, blue circles and red triangles highlight the $2\sigma$ outliers with the smallest $2.3\%$ and largest $2.3\%$ values of $c_{\rm max}/C_{\rm max}$. In the middle row, blue circles highlight the $2\sigma$ outliers with the smallest eccentricities while red triangles highlight the $2\sigma$ outliers with the largest eccentricities.}
\label{fig:deltam} \vspace{0.2cm}
\end{figure*}

A common assumption in modeling tidal mass stripping is that subhalos of a given mass lose a certain fraction of their mass in one dynamical time. For example, \citet{vandenbosch/etal:2005} developed a model in which the mass loss rate of a given subhalo is $\dot{m}\propto m^{1+\zeta}/\tau_{\rm dyn}(z)$ where $\tau_{\rm dyn}(z)\propto(\Delta \rho_{\rm crit}(z))^{-1/2}$ is proportional to the free fall time of a halo, independent of mass. Recently, \citet{jiang/vandenbosch:2014} used numerical simulations to fit the mass dependence, finding $\zeta = 0.07$. This is very close to the case $\zeta=0$, in which the fractional mass loss rate is independent of mass. 

Modeling dynamical friction and sinking due to the resultant loss of angular momentum plays a prominent role in modeling subhalo orbital evolution. The trajectory of subhalos through the host must be modeled accurately. Assumptions typically involve spherical symmetry, wherein the torque of dynamical friction is in the direction of the subhalo orbital angular momentum, and subhalos orbit in the same plane. 

The question naturally arises whether common assumptions such as those discussed above hold for the low-mass subhalos considered here. In the following sections, we examine separately the mass lost per orbit per halo mass, the orbital period per host dynamical time, and the alignment of tidal torques and angular momentum.

\subsubsection{Tidal mass loss: ${\Delta}m/m$}
\label{subsec:resmloss} 

The general picture of mass loss is related to the processes of dynamical friction and tidal stripping, as follows. The continued force of dynamical friction causes an infalling subhalo to slowly descend into its host. As the orbital radius shrinks, so too does the tidal radius, causing the subhalo to continually shed mass from the outside-in. The internal structure of the subhalo is also affected, generally puffing outwards due to the injection of tidal heat, promoting additional mass loss. Mass loss will vary over the course of an orbital period, being strongest (weakest) at pericenter (apocenter) when tidal interactions with the host are greatest (smallest). Tidal mass loss is clearly a complicated process that will depend on both the internal structure of a subhalo as well as its orbital parameters.

We begin our investigation of mass loss in Figure \ref{fig:deltamdist} where we plot distributions in $\Delta m/m_{\rm infall}$. Here, the mass change is $\Delta m = m_{\rm apo,1} - m_{\rm apo,2}$ where $m_{\rm apo,1}$ and $m_{\rm apo,2}$ are the mass at the start and end of the orbit, respectively. The black histogram shows mass loss over the course of the first orbit after infall while the blue histogram shows mass loss over the last orbit before $z = 0$. Recall that we define an orbit to correspond to the time between successive apocenter passages so the ``first'' orbit does not start exactly at infall and the ``last'' orbit does not end exactly at $z = 0$. More specifically, the first orbit begins at the first apocenter after infall\footnote{We find that infall is roughly symmetric about the turnaround point where the subhalo first detaches from the Hubble flow and begins its descent towards the host. In particular, 52\% of subhalos start to lose mass before turnaround while 48\% begin losing mass after turnaround. Hence, for roughly half of the cases, the first orbit begins at the turnaround radius, corresponding to the first apocentre.} while the last orbit terminates at the last apocenter before the present time. We find that 3966 ($65\%$) subhalos finish at least one orbit after infall while 2714 ($44\%$) finish at least two. Since we are interested in comparing how mass loss changes with time, we plot only those 2714 subhalos for which the first and last orbit is different.

Comparing the two distributions in Figure \ref{fig:deltamdist} shows that subhalos tend to lose a larger fraction of their initial mass during their first orbit compared to their last orbit. In particular, the median mass loss in the first orbit is $16\%$ of the initial mass while the median mass loss in the last orbit is about an order of magnitude smaller, at $3\%$ of the initial mass. Note that not all subhalos lose mass over the course of an orbital period. In particular, for both the first and last orbit, roughly $5\%$ of subhalos actually {\em gain} mass. This likely occurs either through direct merger with smaller systems or, more gradually, through the accretion of surrounding material. 

We proceed to investigate the dependence of mass loss on subhalo properties. The top row of Figure \ref{fig:deltam} shows mass loss versus mass ratio at the start of the orbit. For both orbits, more massive subhalos tend to lose more mass on average. Normally, we would expect this result on the basis of a dynamical friction argument whereby the oribts of massive subhalos are preferentially dragged into the depths of the host, promoting enhanced mass loss. However, we do not expect this argument to apply here since the dynamical friction merging timescale for $\mu \lesssim 10^{-3}$ subhalos is much longer than the Hubble time \citep{boylan-kolchin/etal:2008}. 

Instead, the observed correlation with mass is the result of the mass-concentration relation which states that more massive subhalos will be less concentrated on average. We plot as blue circles (red triangles) the $2\sigma$ outliers with the smallest $2.3\%$ (largest $2.3\%$) values of $c_{\rm max}/C_{\rm max}$ at the start of the orbit. From the definition of $c_{\rm max}$ in equation (\ref{eq:cproxy}), the ratio $c_{\rm max}/C_{\rm max}$ describes the relative central density of the subhalo to the host. In both panels a clear dichotomy emerges with the least (most) concentrated, and most (least) massive, subhalos loosing (retaining) more mass per obit. 

This is made more apparent in the middle row of Figure \ref{fig:deltam} where we see a strong negative slope in mass loss versus concentration. There is still considerable scatter at fixed concentration which can be partly attributed to eccentricity. Comparing blue circles and red triangles shows that for fixed concentration, more circular (radial) orbits tend to retain (lose) more mass on average. A direct comparison is plotted in the bottom row of Figure \ref{fig:deltam} where we detect a small correlation between mass loss and eccentricity. 

We have also checked for correlation between mass loss and pericenter, $r_{\rm peri}$. One would expect that subhalos plunging further into the depths of the host, where tidal forces are strong, would experience enhanced mass loss. We instead find almost no correlation with $r_{\rm peri}$\footnote{Note that our calculation of $r_{\rm peri}$ is based upon a cubic spline interpolation of subhalo radial distance from discrete VL2 snapshots (see \S\ref{sec:vl2data}). The snapshots are separated by 0.688 Gyr which may lead to a potentially crude estimation of the true $r_{\rm peri}$. Another way to estimate $r_{\rm peri}$ is to solve the roots in the equation of motion of a point particle of energy $E$ and angular momentum $L$ in a static NFW host potential \citep[see, e.g.,][]{binney/tremaine:1987}. This may lead to a better determination of $r_{\rm peri}$ as $E$ and $L$ vary more smoothly with time than radial distance. Nevertheless, we have checked that using $r_{\rm peri}$ computed in this way changes neither the results on mass loss versus eccentricity nor mass loss versus pericentre.}. The reason is that subhalos closer to the host center tend to be more concentrated (see \S\ref{sec:results-rinf}) which washes out the dependence on $r_{\rm peri}$. 

We conclude that tidal mass loss in the regime of low-mass subhalos is most directly correlated with concentration. When concentration is held fixed, we find no trend in mass loss with varying mass. The apparent trend seen when comparing mass loss versus mass is simply a reflection of the fact that mass is correlated with concentration. This result makes physical sense in the limit of weak dynamical friction since it is the density of a subhalo, relative to its host, that determines how tightly a subhalo on a stable orbit retains its contents \citep[e.g.,][]{taffoni/etal:2003}. At fixed concentration, subhalos on more eccentric (i.e., radial) orbits tend to lose more mass than subhalos on circular orbits. This may highlight the importance of tidal heating which results when a rapidly varying gravitational potential injects energy into subhalos, puffing them outwards and promoting further mass loss \citep[e.g.,][]{hayashi/etal:2003}.

\begin{figure} \smjustify
\includegraphics[width=\smhwidth]{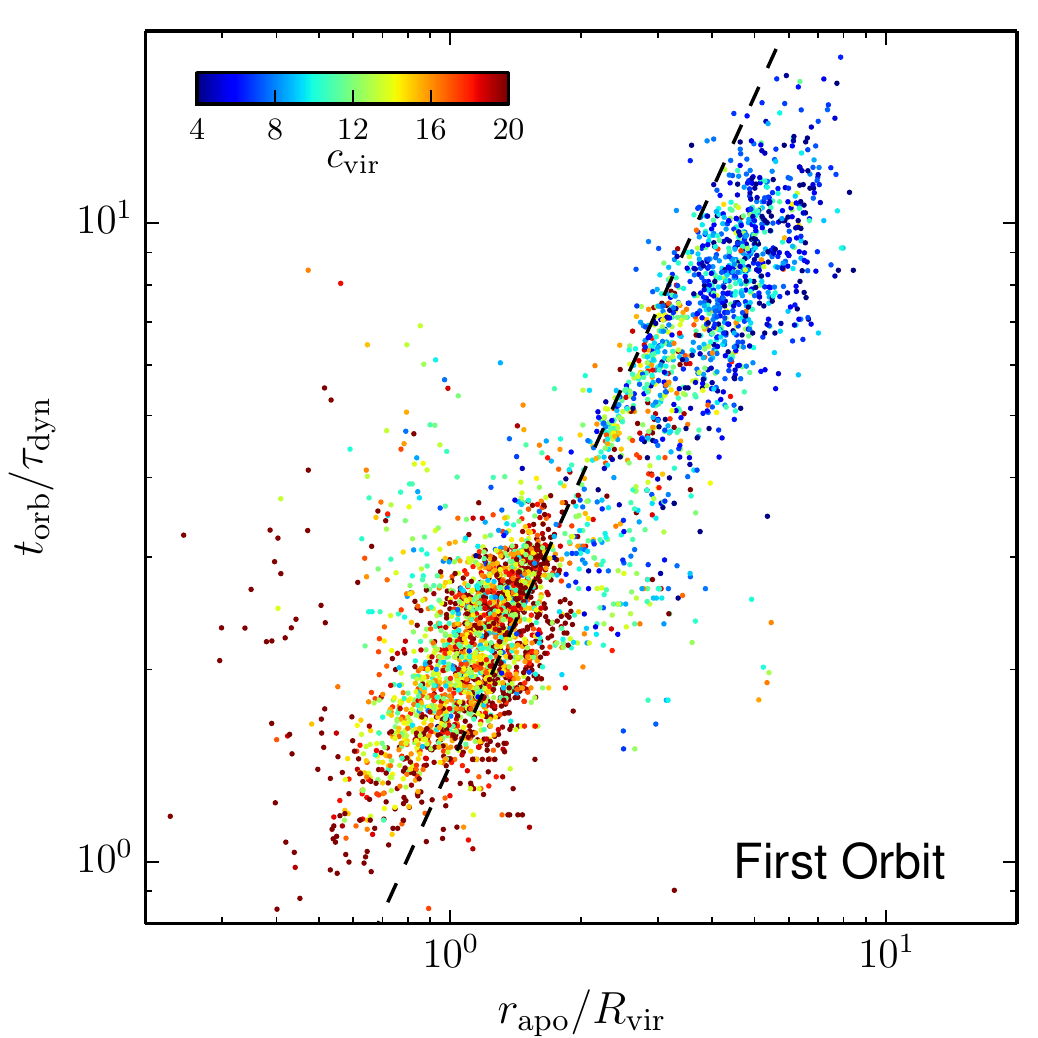}
\caption{Orbital period, $t_{\rm orb}$, in units of the halo dynamical time, $\tau_{\rm dyn}$, for subhalos on their first orbit, colored by concentration, as a function of first apocenter, $r_{\rm apo}$, in units of the host virial radius. All quantities are determined at the beginning of the orbit. Shown as the dashed line is the relationship expected for radial orbits (eccentricity $e=1$) orbiting a point with mass $M_{\rm vir}$. }
\label{fig:tovertau} \vspace{0.2cm}
\end{figure} 

\subsubsection{Orbital period}
\label{subsec:resorbdyn}

The dynamical time for a halo is usually defined as the free-fall time of a test particle in a static, uniform sphere at the virial density, 
\begin{equation}
\tau^2_{\rm dyn}\equiv [16G\rho_{\rm crit}\Delta/(3\pi)]^{-1}=\frac{\pi^2R_{\rm vir}^3}{4GM_{\rm vir}}.
\label{eq:tau}
\end{equation}
It is natural to expect, all else being equal, that timescales within the halo should scale in proportion to this dynamical time. For example, the time to complete an orbit is roughly proportional to the dynamical time. Similar scaling arguments apply for timescales other than orbital period, such as the tidal mass loss time, $m/\dot{m}$. Departures from a simple linear scaling with the dynamical time occur because orbital shapes vary from subhalo to subhalo. For example, halos on larger orbits should have longer orbital times, with a correlation between semi-major axis and period that reflects the mass distribution of the host halo around the virial radius. 

\begin{figure} \smjustify
\includegraphics[width=\smhwidth]{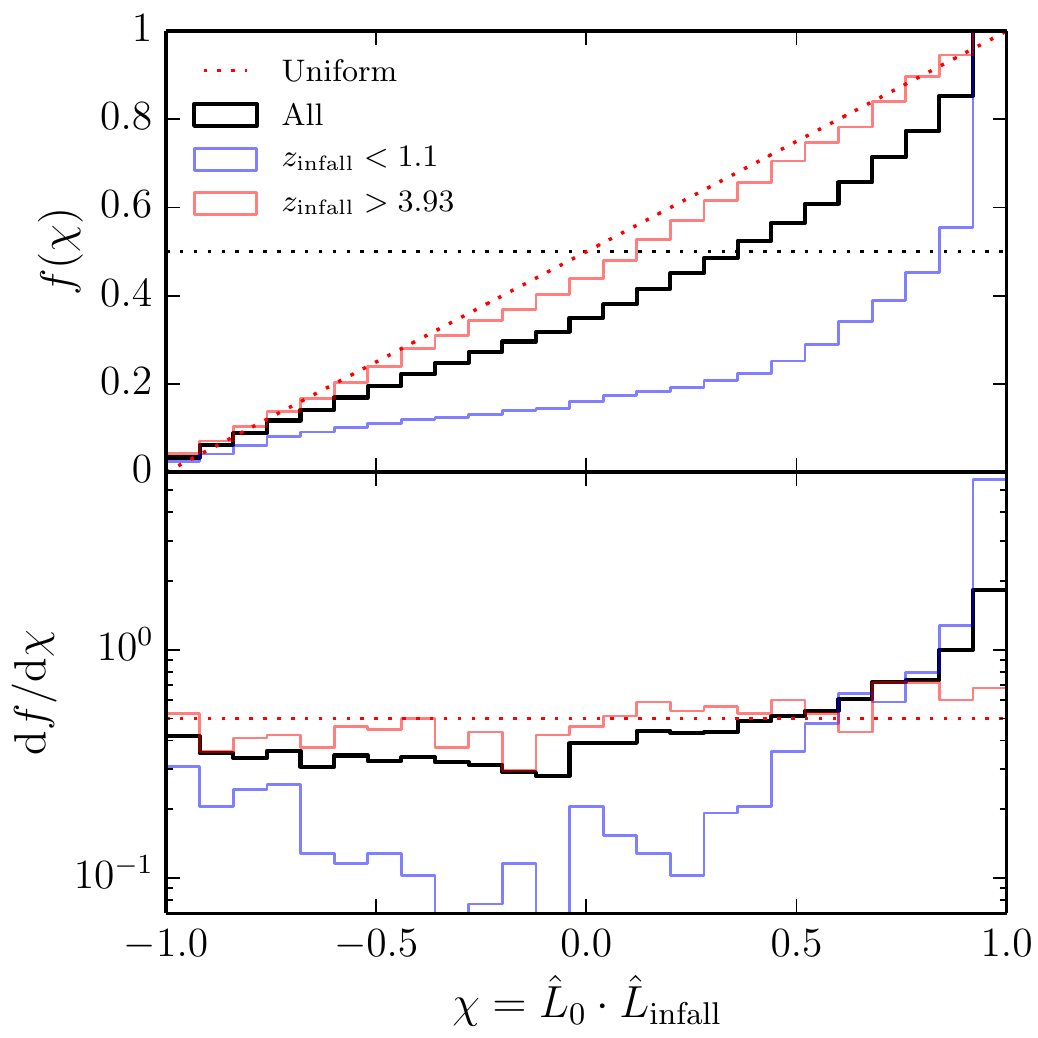}
\caption{Cumulative (top panel) and differential (bottom panel) distributions in the dot product between the angular momentum normal vector at $z = 0$ and infall. The black histogram traces the total sample of subhalos while the blue and red histograms show the $1\sigma$ outliers with latest and earliest infall, respectively. Those subhalos that fell in recently tend to have their present-day angular momentum vector more closely aligned with the infall direction while subhalos that fell in earlier approach a uniform distribution in $\chi$ (shown for comparison as the dotted red line).}
\label{fig:angmom} \vspace{0.2cm}
\end{figure} 

In Figure \ref{fig:tovertau} we show the orbital period of halos just after they fall in, defined as the time between the first two apocenters after infall. The orbital period is plotted in units of the dynamical time, $\tau_{\rm dyn}$, at the beginning of the orbit. On the $x$-axis we show $r_{\rm apo}/R_{\rm vir}$.
The first striking trend is the large spread in $t_{\rm orb}/\tau_{\rm dyn}$ values. There is also a similarly large spread, about an order of magnitude, in the apocenters, $r_{\rm apo}/R_{\rm vir}$ of halos on their first orbit. In fact, there is a strong correlation between orbital period and apocentric distance, as expected. Interestingly, there are two ``clouds" of subhalos. Those with high concentration at small radius, and those with low concentration at large radius. 

Also shown is the relationship expected for radial orbits (where the semi-major axis $a=r_{\rm apo}/2$) around a point mass with $M=M_{\rm vir}$, $t^2_{\rm orb}=4\pi^2a^3/(GM_{\rm vir})=\pi^2r_{\rm apo}^3/(2GM_{\rm vir})$. Combining with equation (\ref{eq:tau}), we obtain 
\begin{equation}
\frac{t_{\rm orb}}{\tau_{\rm dyn}} = 4\left(\frac{r_{\rm apo}}{2R_{\rm vir}}\right)^{3/2}.
\end{equation}
Subhalos would lie along this line only if they were on radial orbits and all the halo mass was located at the halo center. In general, departures from radial orbits ($a>r_{\rm apo}/2$) lead to longer orbital times, while the presence of matter outside the virial radius (i.e. the overdensity associated with continuous infall onto the host) leads to shorter orbital times. The latter effect could be responsible for the shorter times at $r_{\rm apo}>2R_{\rm vir}$, although more information about the evolving density profile outside the virial radius would be required to make a quantitative comparison.

\subsubsection{Orbital plane}
\label{subsec:evcirc}

A common assumption made in models of substructure evolution is spherical symmetry of the host. Subhalo orbits are generally integrated in either a static potential or one that dynamically adjusts (e.g., through mass accretion) in a spherically symmetric manner. In either case, the direction of the orbital angular momentum vector is conserved since it is aligned with the direction of the torque. Hence, an obvious test of spherical symmetry within VL2 is to look for changes in the orientation of the orbital plane.

In Figure \ref{fig:angmom} we plot the distribution in the dot product between the angular momentum normal vector at $z = 0$ and infall:
\bq
\chi = \hat{L}_0 \cdot \hat{L}_{\rm infall}.
\label{eq:angmom}
\eq
The black histogram shows the distribution for all subhalos while the blue and red histograms show the $1\sigma$ outliers with the latest and earliest infall, respectively. The median $\chi$ for all subhalos is 0.39 while subhalos with the earliest and latest infall time have median values 0.16 and 0.88, respectively. There is a clear trend of recently infalling halos remaining in the same orbital plane while subhalos with early infall have their orientation randomly aligned. 

Subhalo orbits are continuously torqued after infall, in a direction that is not aligned with the angular momentum vector. Subhalos spending more time in the host experience larger changes in $\hat{L}$. $98\%$ of subhalos with $z_{\rm infall} < 1.1$ do not finish an orbital period by $z = 0$ while $65\%$ of subhalos with $z_{\rm infall} > 3.93$ finish at least three orbits. The latter population approach a uniform distribution in $\chi$, indicating that memory of the initial orbital plane is lost after a few orbits within the host. It is clear that the assumption of spherical symmetry does not apply.

This result is not too surprising, however, since dark matter halos are generally triaxial in shape and the host will experience anisotropic mass redistribution as massive objects are biased toward filamentary accretion \citep[see also, e.g.,][]{zemp/etal:2009}. Another possible source of orbital torque is substructure interaction. \citet{slater/bell:2013} used VL2 to show that a significant fraction of subhalos accrete as groups with correlated trajectories that lead to frequent interaction over time. Such interactions can lead to a complex redistribution of orbital energy and angular momenta for the low-mass subhalos that are abundant here \citep{sales/etal:2007,ludlow/etal:2009}. Though a more detailed inspection of orbits is required to assess the significance of these effects in VL2, our result highlights the importance of considering host anisotropy and subhalo interaction in semi-analytic models of substructure evolution. 

\begin{figure}[!t] \smjustify
\includegraphics[width=\smhwidth]{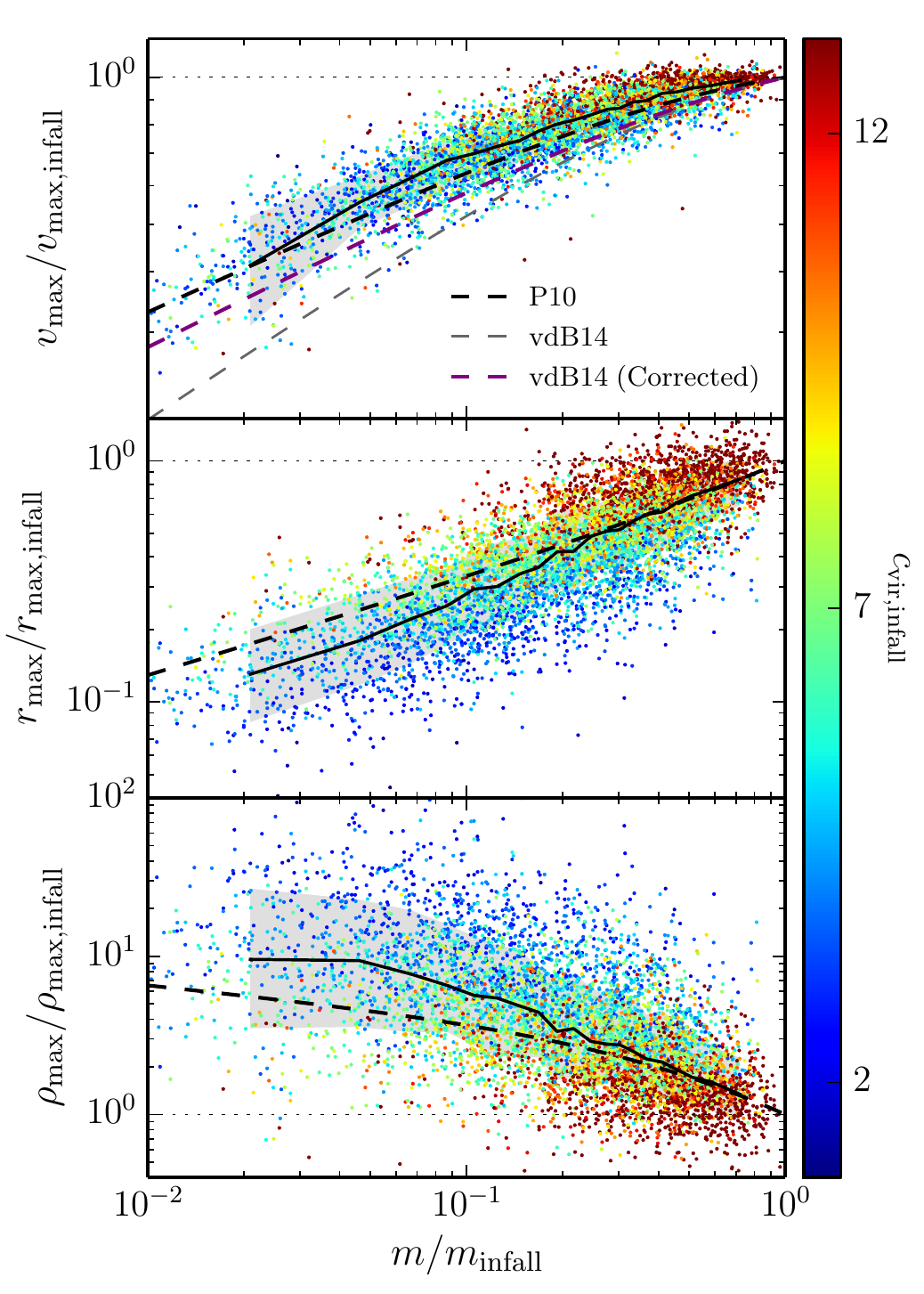}
\caption{Evolution of $\vmx$ (top), $\rmx$ (middle), and $\pmx$ (bottom) as a function of mass retained at $z = 0$. In each panel, points represent individual subhalos coloured according to concentration at infall while the solid black line shows the median trend with the associated $1\sigma$ scatter shaded in gray. The P10 relation for isolated NFW subhalos of fixed concentration falling into a static host is shown as the dashed black line in each panel. The thin dashed gray line in the top panel traces the Bolshoi $\vmx$ relation as reported in \citet{vandenbosch/jiang:2015} while the dashed purple line traces the corrected relation including only sufficiently resolved subhalos in Bolshoi (van den Bosch, private communication).}
\label{fig:rmaxev} \vspace{0.2cm}
\end{figure}

\subsubsection{Subhalo internal structure: $\rmx$ and $\vmx$}

The main observable properties of luminous subhalos are their velocity structure, often described in terms of the circular velocity profile, $v^2=GM(<r)/r$. In particular, most dynamical measurements provide robust constraints on the maximum circular velocity, $\vmx$, and the radius at which this occurs $\rmx$. In this section we show evolution in these two quantities as subhalos descend into the host. We refer to this as evolution in internal structure in the sense that $\vmx$ and $\rmx$ describe central density with $\rho_{\rm max} \propto (v_{\rm max}/r_{\rm max})^2$ being the mean density within $\rmx$.

A number of previous works \citep{hayashi/etal:2003,penarrubia/etal:2008,penarrubia/etal:2010} have studied the evolution in $\rmx$ and $\vmx$ using numerical simulations where isolated subhalos are dropped into the potential of a static host. These studies come to the same conclusion that $\vmx$ and $\rmx$ evolve along tightly defined trajectories when written in terms of the mass fraction retained after infall. In particular, defining $x = m/m_{\rm infall}$ and taking $y$ to represent either $r_{\rm max}/r_{\rm max,infall}$ or $v_{\rm max}/v_{\rm max,infall}$, it is found that subhalos starting at $(x, y) = (1,1)$ move steadily along the track
\bq
y(x) = \frac{2^\alpha x^\beta}{(1+x)^\alpha},
\label{eq:penfit}
\eq
where $\alpha$ and $\beta$ are fitting coefficients. For the case of NFW subhalos, \citet[][P10]{penarrubia/etal:2010} find the result $(\alpha, \beta) = (0.4, 0.3)$ for $\vmx$ and $(\alpha, \beta) = (-0.3, 0.4)$ for $\rmx$.

\begin{figure*}
\begin{center}
\includegraphics[width=0.9\textwidth]{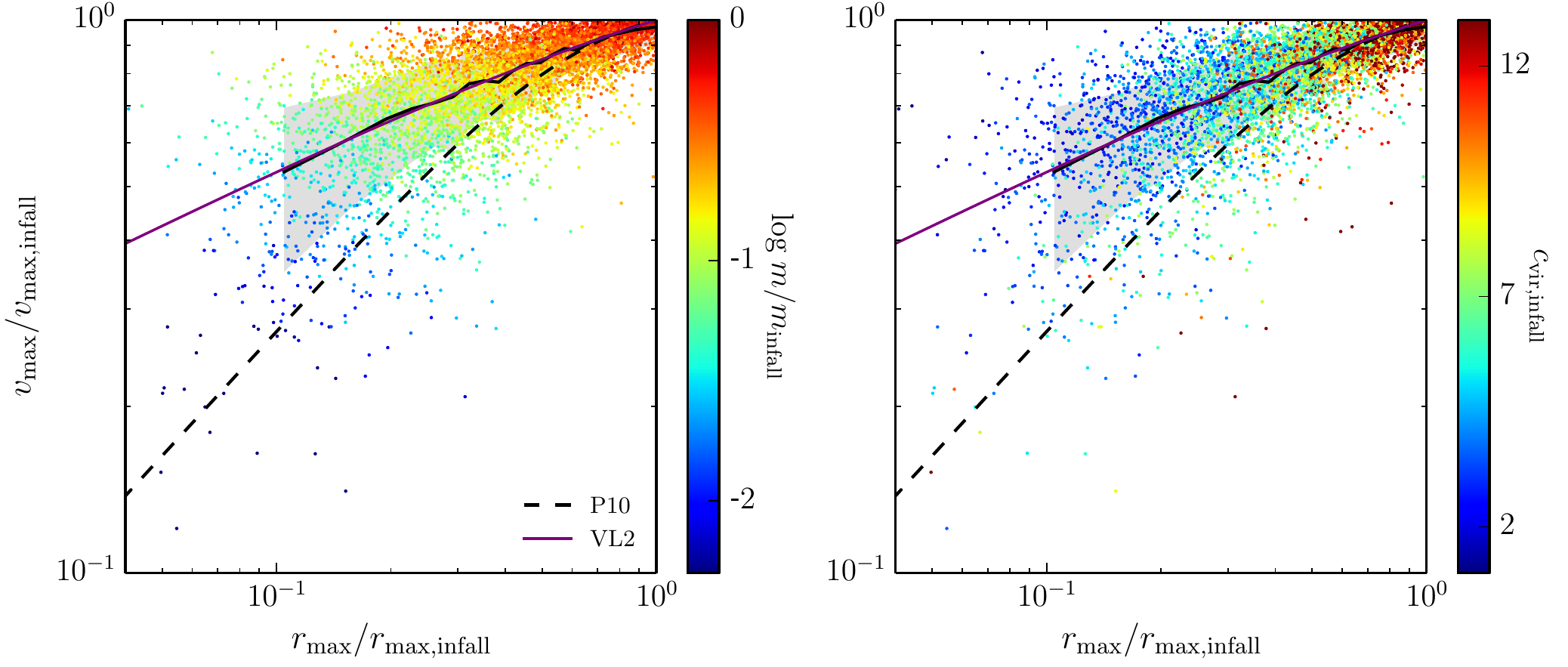}
\end{center}
\caption{Evolution of subhalos in the normalized $v_{\rm max}-r_{\rm max}$ plane. Left panel shows subhalos coloured according to the fraction of mass remaining at $z = 0$ while the right panel colours subhalos according to their concentration at infall. The thick black line shows the median trend with the associated $1\sigma$ scatter shaded in gray. The thin purple line shows equation (\ref{eq:penfit}) with $(\alpha, \beta) = (0.25, 0.34)$ chosen to match the median relation. The dashed black line shows the P10 relation for isolated NFW subhalos of fixed concentration falling into a static host.}
\label{fig:rvmaxev} \vspace{0.5cm}
\end{figure*}

In Figure \ref{fig:rmaxev} we plot the ratio of the present-day values of $\vmx$, $\rmx$, and $\pmx$ to their infall values versus the fraction of mass retained at $z=0$. In each panel, the solid black line traces the median relation and the shaded region shows the $1\sigma$ scatter. The vast majority of subhalos experience reduction in $\rmx$ and $\vmx$, with a larger suppression in the former, leading to a net increase in $\pmx$ with increasing mass loss.  The median relation in each panel can be compared to the dashed black line showing the P10 result. We find VL2 agrees well with P10 for $\vmx$ but begins to diverge at low mass retention for $\rmx$ and $\pmx$. The dashed gray line in the top panel shows equation (\ref{eq:penfit}) with $(\alpha, \beta) = (0.60, 0.44)$ which was reported by \citet{vandenbosch/jiang:2015} to fit evolution in $\vmx$ for subhalos in the Bolshoi simulation. The VL2 data sits systematically above the Bolshoi result. The dashed purple line shows a corrected form $(\alpha, \beta) = (0.36, 0.33)$ which fits the Bolshoi relation when insufficiently resolved subhalos are removed from the sample (van den Bosch, private communication). This shows much better agreement with the VL2 result. 

The points in Figure \ref{fig:rmaxev} are coloured in terms of subhalo concentration at infall. In the case of $\vmx$ we do not see much dependence on concentration other than the fact that subhalos with larger $c_{\rm vir}$ tend to have fallen in more recently and therefore have not had as much time to evolve to the left side of the plot. In contrast, $\rmx$ and $\pmx$ show strong stratification in $c_{\rm vir}$ with the least concentrated subhalos showing systematically greater reduction in $\rmx$ and enhancement in $\pmx$. We offer a heuristic explanation as follows. Subhalos on slowly sinking orbits experience mass loss until the tidal radius shrinks to the point at which the mean interior density is proportional to the local density of the host. Since subhalos are exposed to (roughly) the same local density, those that were initially more dense (i.e., larger $c_{\rm vir}$) naturally approach a smaller value of $\pmx/\rho_{\rm max,infall}$ at late times. 

P10 do not find significant scatter since they consider subhalos of fixed concentration. They do find, however, that varying the {\em shape} of the subhalo inner density profiles at fixed concentration changes the coefficients $\alpha$ and $\beta$. We find the complementary result that varying concentration at fixed shape leads to substantially different structural evolution. 

This point is made more illuminating by plotting $\vmx$ versus $\rmx$, as in Figure \ref{fig:rvmaxev}. In the left panel we colour points according to the mass retained at $z = 0$ while the right panel shows concentration at infall. The left panel shows a clear gradient in colour, reinforcing the notion of previous works \citep{hayashi/etal:2003,penarrubia/etal:2008,penarrubia/etal:2010} that evolution in internal structure does not depend on {\em how} mass is lost, but only {\em how much} mass is lost. As expected, however, subhalos are not bound to a single trajectory in the $\vmx-\rmx$ plane. The right panel shows that scatter at fixed mass loss can be attributed to concentration, as in Figure \ref{fig:rmaxev}. The median evolution for all subhalos is shown as the solid black line in each panel with $1\sigma$ scatter shaded in gray. The purple line fits the median trend using equation (\ref{eq:penfit}) with $(\alpha, \beta) = (0.25, 0.34)$. This sits above the P10 result that was derived from subhalos of fixed $c_{\rm vir} = 23$; a considerably larger value than the median concentration of 7 found in VL2. 


\section{Disrupted Subhalo Population}
\label{sec:dispop}

\begin{figure*}
\begin{center}
\includegraphics[width=0.9\textwidth]{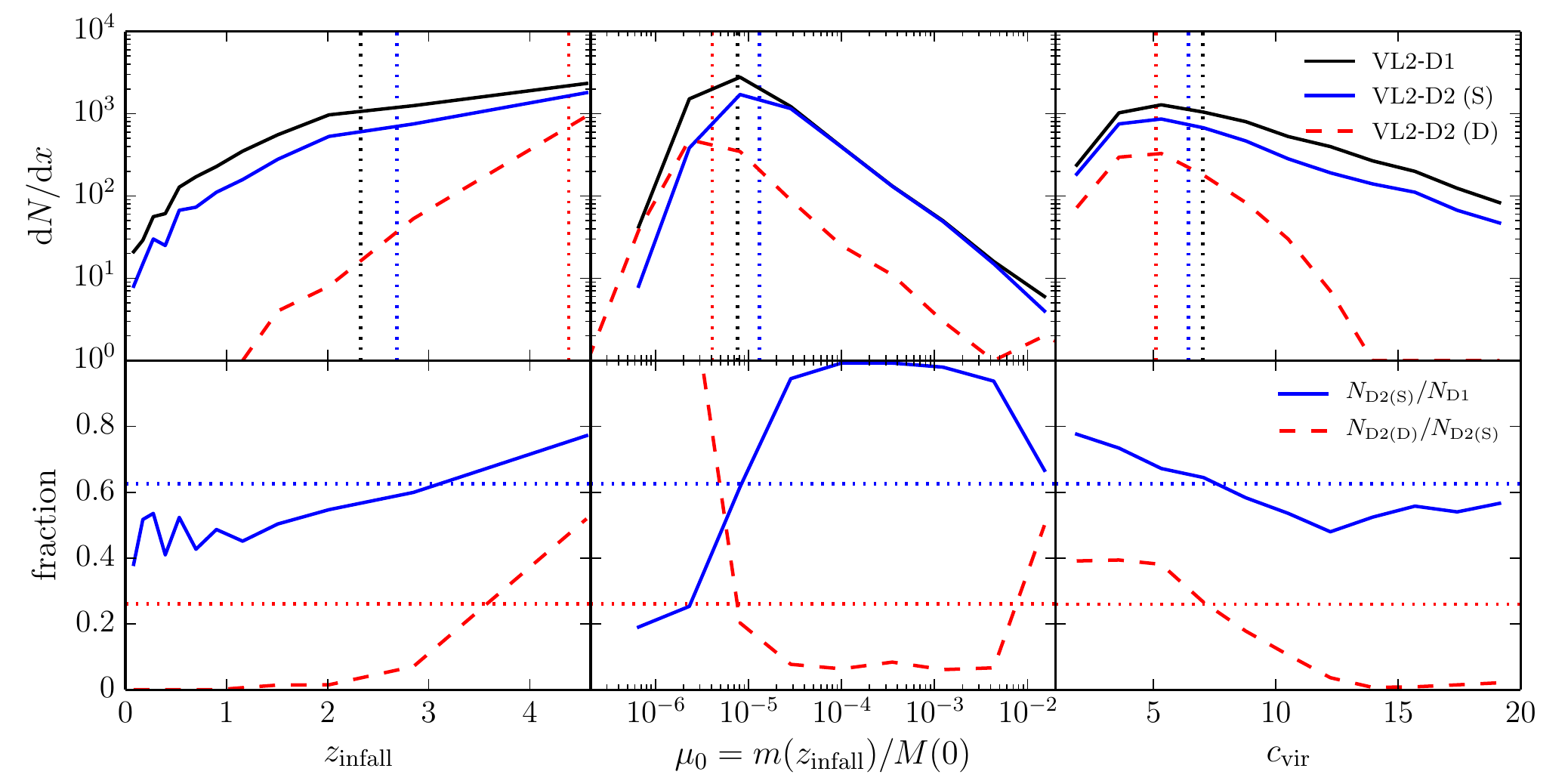}
\end{center}
\caption{Top panels show the distribution of subhalos based on $z_{\rm infall}$ (left), mass ratio at infall (middle), and concentration at infall (right). The solid black line shows the distribution of subhalos from the main catalogue, VL2-D1, considered in the preceding sections of this paper. The solid blue (dashed red) line traces the distribution of surviving (disrupted) subhalos from VL2-D2. The vertical dotted lines show the median value of the distribution of the corresponding colour. The dashed red lines in the bottom panels trace the ratio $N_{{\rm D2} ({\rm D})}/N_{{\rm D2} ({\rm S})}$ of disrupted to surviving subhalos in VL2-D2 with the dotted red line showing the mean value $1000/3843$. The solid blue lines trace the fraction $N_{{\rm D2} ({\rm S})}/N_{{\rm D1}}$ of surviving subhalos in VL2-D2 to surviving subhalos in VL2-D1. The dotted blue line shows the mean fraction $3843/6145$.}
\label{fig:fdisrpt} \vspace{0.5cm}
\end{figure*}

Up to this point we have only considered the population of subhalos that {\em survive} to the present day. This leaves open questions regarding any potential biases that may exist in our results due to the omission of {\em disrupted} subhalos. In this section we attempt to address these questions by making use of the second VL2 public catalogue. 

The second catalogue contains a similar set of evolutionary tracks as the main catalogue except that it pertains to the 20000 largest systems in the simulation box identified at $z = 4.56$. The two catalogues are not mutually exclusive as some of the surviving subhalos at $z = 0$ also happened to be of the largest systems present at $z = 4.56$. The utility of the second catalogue is that it contains subhalos that disrupt prior to the present day. Though this does not constitute the {\em full} ensemble of disrupted subhalos (some subhalos that disrupt were not of the largest systems at $z = 4.56$) it should be enough to elucidate differences between surviving and disrupted subhalos. For convenience we henceforth refer to the main VL2 data set (considered in all preceding sections) as VL2-D1 and refer to the second data set as VL2-D2.

We apply the same framework outlined in \S\ref{sec:vl2data} to VL2-D2. Namely, we identify subhalos as those systems that at some point passed within the instantaneous virial radius of the host. Of these subhalos we identify the surviving population as those that still exist as intact objects at $z = 0$. Conversely, the disrupted population consists of those subhalos that fall below the mass resolution of the VL2 halo finder some time before reaching $z = 0$. We find a total of 4843 subhalos in VL2-D2 of which 3843 (79\%) belong to the surviving group and 1000 (21\%) belong to the disrupted group. Note that $\sim63\%$ of the 6145 subhalos from VL2-D1 are also part of VL2-D2.

We examine in Figure \ref{fig:fdisrpt} the dependence of survivability on infall redshift, mass ratio, and concentration. In the top panels, we plot the distribution in each quantity for all surviving (disrupted) subhalos from VL2-D2 as solid blue (dashed red) lines. For comparison, the solid black line traces the distribution in each quantity for the VL2-D1 subhalos. The vertical dotted lines denote the median value of the distribution with the corresponding colour. In the bottom panels, the dashed red line traces the ratio,  $N_{{\rm D2} ({\rm D})}/N_{{\rm D2} ({\rm S})}$, of disrupted to surviving subhalos in VL2-D2, with the horizontal red line denoting the mean value $1000/3843$. To test for bias in the surviving fraction of VL2-D2 subhalos, the solid blue line traces the ratio, $N_{{\rm D2} ({\rm S})}/N_{{\rm D1}}$, of the surviving subhalos in VL2-D2 to the full ensemble of surviving subhalos contained in VL2-D1. The horizontal blue line denotes the mean fraction $3843/6145$.

We begin with infall redshift. The blue line in the bottom left panel shows that VL2-D2 is slightly biased towards containing those surviving subhalos with larger values of $z_{\rm infall}$. This reflects the fact that subhalos with smaller values of $z_{\rm infall}$ were less likely to exist as large objects at $z = 4.56$ when VL2-D2 was constructed. Comparing the surviving and disrupted populations in VL2-D2 shows that subhalos with $z_{\rm infall} \gtrsim 3$ are much more likely to belong to the latter group. That is, subhalos spending more time exposed to the tidal field of the host are more likely to disrupt by the present day.

Next we investigate mass ratio at infall. The middle panels of Figure \ref{fig:fdisrpt} show little dependence of survivability in the mass range $10^{-5} \lesssim \mu_0 \lesssim 10^{-2}$. For larger masses, we expect the disrupted fraction to increase as dynamical friction preferentially causes massive subhalos to plummet into the depths of the host where tidal forces are strongest. We indeed see an upturn at $\mu_0 \gtrsim 10^{-2}$ though VL2 is hindered by small number statistics in this regime to make a meaningful statement here. We also observe a rapid rise in the disrupted fraction for $\mu_0 \lesssim 10^{-5}$. This is expected due to the finite mass resolution of the simulation -- subhalos closer to the resolution limit are more likely to ``disrupt''. There is a related drop in the blue line in the bottom middle panel indicating that the missing surviving population in VL2-D2 are almost exclusively low-mass subhalos. This reflects the early infall bias in VL2-D2: subhalos nearer the resolution limit are only likely to survive to $z = 0$ if they infall later. 

We finish by examining infall concentration in the right panels of Figure \ref{fig:fdisrpt}. First, we see that VL2-D2 is biased towards containing low-concentration surviving subhalos, consistent with the early infall bias via the concentration-redshift relation (see Figure \ref{fig:crvz}). Moreover, of all subhalos in VL2-D2, those with $c_{\rm vir} \lesssim 5$ are much more likely to disrupt by $z = 0$. This is consistent with our previous finding that low-concentration subhalos are more susceptible to tidal stripping from the host (see Figure \ref{fig:deltam}). In fact, the strong dependence of survivability on $c_{\rm vir}$ and not on $\mu_0$ strengthens the notion that mass loss is more strongly connected to concentration than mass for low-mass ($\mu_0 \lesssim 10^{-3}$) subhalos. 

\begin{figure} \smjustify
\includegraphics[width=\smhwidth]{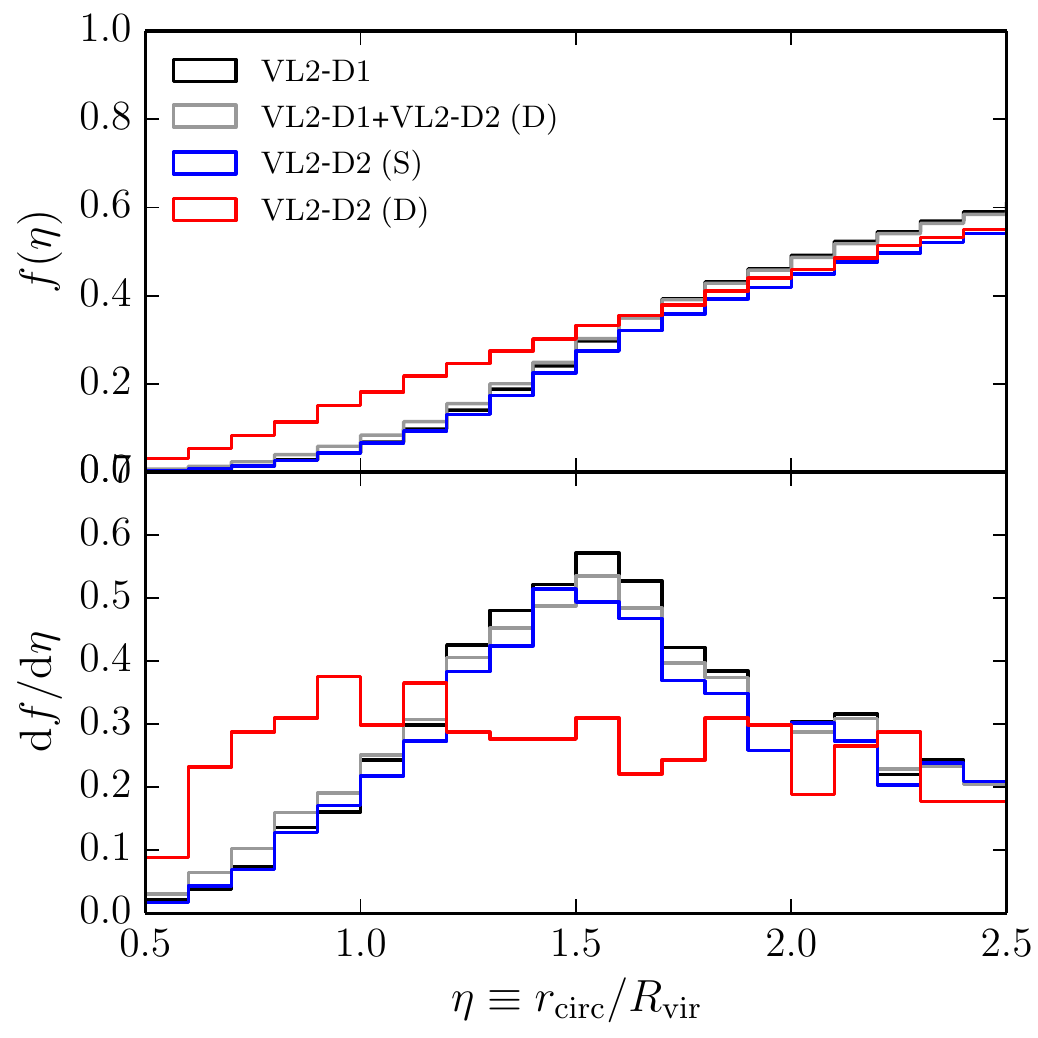}
\caption{Cumulative (top panel) and differential (bottom panel) distributions in $\eta$ for all surviving VL2 subhalos (black histogram), the surviving subhalos in the VL2-D2 catalogue (blue histogram), and the disrupted subhalos in the VL2-D2 catalogue (red histogram). The gray histogram shows the result of combining the total population of surviving subhalos with the VL2-D2 disrupted population. In each case we report $\eta$ at $z_{\rm infall}$ for an NFW host potential. The results are qualitatively similar if we consider virial crossing with a point mass host potential.}
\label{fig:eta-D2} \vspace{0.2cm}
\end{figure}

We now shift focus to the infall distribution of orbital energy and angular momentum for surviving versus disrupted subhalos. In Figure \ref{fig:eta-D2} we plot the infall distribution of $\eta$ for the surviving (disrupted) subhalos in VL2-D2 as a blue (red) histogram. The blue histogram can be compared to the black histogram showing the infall distribution for all surviving subhalos in VL2-D1 (i.e., the black histogram in Figure \ref{fig:eta}). We see good agreement between the surviving VL2-D2 subhalos and the full VL2-D1 population. The disrupted population, on the other hand, agrees well with the other curves for $\eta > 2$, but displays a much flatter distribution for smaller $\eta$. Most importantly is the excess at $\eta < 1$, indicating that subhalos strongly bound to the host at infall are preferentially disrupted by $z = 0$. 

This result suggests that the black histogram is suppressed at $\eta < 1$ compared to the distribution of {\em all} subhalos that ever fell onto the host, regardless of survivability. The gray histogram in Figure \ref{fig:eta-D2} shows the result of combining the disrupted subhalos in VL2-D2 with the VL2-D1 catalogue. Doing so results in only a minor change to the black histogram since there are six times fewer disrupted subhalos than surviving subhalos. Though we do not have access to the full ensemble of disrupted subhalos, it seems unlikely that we are missing a large enough fraction for there to be a significant impact on the infall distribution and fitting functions for $\eta$ presented in \S\ref{sec:results-etainf}.

\begin{figure} \smjustify
\includegraphics[width=\smhwidth]{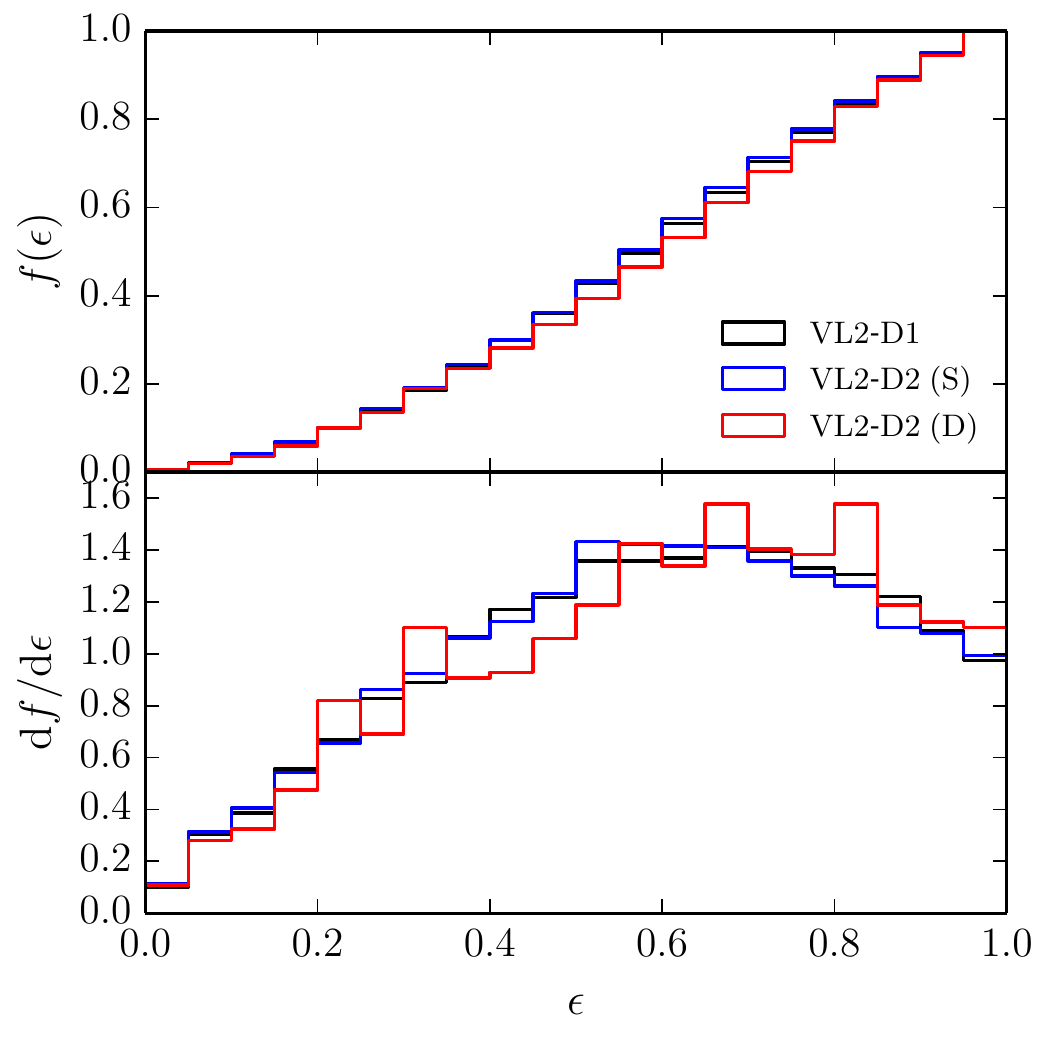}
\caption{Cumulative (top panel) and differential (bottom panel) distributions in $\epsilon$ for all surviving VL2 subhalos (black histogram), the surviving subhalos in the VL2-D2 catalogue (blue histogram), and the disrupted subhalos in the VL2-D2 catalogue (red histogram). In each case we report $\epsilon$ at $z_{\rm infall}$ for an NFW host potential. The results are qualitatively similar if we consider virial crossing with a point mass host potential.}
\label{fig:circ-D2} \vspace{0.2cm}
\end{figure}

In Figure \ref{fig:circ-D2} we plot the infall distribution of $\epsilon$ for the surviving (disrupted) subhalos in VL2-D2 as a blue (red) histogram. The surviving population can be compared to the black histogram showing the infall distribution for all surviving subhalos in VL2-D1 (i.e., the black histogram in Figure \ref{fig:circ}). In this case, we see good agreement between all histograms. Hence, unlike orbital energy, angular momentum does not appear to play a significant role in determining the survivability of subhalos. The infall distribution and fitting functions for $\epsilon$ presented in \S\ref{sec:results-epsinf} are therefore robust to the inclusion of disrupted subhalos. 

\section{Summary}
\label{sec:summary}

We have analyzed the publicly available VL2 halo catalogue in order to characterize the infall properties, orbital dynamics, and structural evolution of over 6000 subhalos within a galactic host. Our main focus is on the role of subhalo concentration in each of these categories, and how this relates to $z = 0$ observables, such as circular velocity and size. 

We define subhalo infall as the time when a halo reaches maximum mass. In other words, a halo becomes a subhalo when its growth is halted, mainly due to tidal truncation from the host. After infall, subhalos experience mass loss from tidal forces and exhibit internal readjustment as they gradually sink toward the host center. We focus on low-mass subhalos for which dynamical friction plays only a minor role, leading to qualitatively different behaviour than is often described for high-mass subhalos. In the following paragraphs we report the main results of our paper.

\vspace{\bulletskip}
{\bf Subhalo mass function:} We compare the unevolved and evolved subhalo mass functions with the results published from the Aquarius simulations. The unevolved mass function uses the mass of each subhalo {\em when it fell in} while the evolved mass function uses the mass at $z = 0$, showing the cumulative effect of tidal mass loss after infall. While the unevolved mass functions agree well, the evolved VL2 mass function is systematically lower, by $30\%$, corresponding to a downward shift in mass (Figure \ref{fig:dndm}). The lower normalization of the VL2 simulation ($\sigma_8=0.74$ vs. $\sigma_8=0.9$) could be the origin, although the physical explanation remains unclear (see \S\ref{sec:results2}).

\vspace{\bulletskip}
{\bf Properties at infall:} Several important relationships among subhalo properties at the time of infall emerge: (1) The typical infalling halo mass does not evolve significantly with time, with a value of $\sim 10^7\ M_\odot$ (Figure \ref{fig:crvz} and also Figure \ref{fig:fdisrpt}). (2) Rare, massive halos fall in much later than less massive ones, as expected in hierarchical structure formation (Figure \ref{fig:zinfall}).
(3) Halos that fall in earlier have lower concentrations, consistent with the well-known concentration-mass-redshift relationship for dark matter halos at fixed mass \citep[e.g.,][see Figure \ref{fig:crvz}]{klypin/etal:2011}. (4) Halos that fall in earlier or, equivalently, that have lower concentrations at infall, tend to experience tidal growth truncation at a larger radius (Figures \ref{fig:rinfall} and \ref{fig:crvz}).

\vspace{\bulletskip}
{\bf Energy and angular momentum:} The orbital energy and angular momentum of subhalos at infall are significantly different than reported in previous studies that focused on more massive subhalos. The low-mass subhalos found in VL2 are skewed toward lower specific binding energy (Figure \ref{fig:eta}) and slightly more circular orbits (Figure \ref{fig:circ}). This is consistent with the recent results of \citet{jiang/etal:2014b}. The explanation for these trends may be related to the environment in which objects form -- massive subhalos fall into the host preferentially along filaments, leading to tightly bound, radial orbits -- and/or gravitational interactions prior to infall that preferentially inject energy and tangential motion into low-mass subhalo orbits.

\vspace{\bulletskip}
{\bf Mass loss:} We find that subhalos undergo most of their mass loss on the first pericenter passage, with a median mass loss fraction of $\sim 0.2$  (Figure \ref{fig:deltamdist}). The fraction of mass lost in the first pericenter passage is most correlated with halo concentration at infall -- less concentrated halos tend to undergo more mass loss, nearly independent of mass (Figure \ref{fig:deltam}). There is a significant but less pronounced correlation of mass loss fraction with subhalo mass, but this trend is only apparent, being explained by the fact that more massive halos have lower concentrations on average and are thus more susceptible to tidal effects. Scatter in mass loss at fixed concentration can be mainly attributed to eccentricity with radial orbits tending to lose more mass than circular orbits. 

\vspace{\bulletskip}
{\bf Orbital period:} 
The period of the first orbit after infall is roughly proportional to the dynamical time of the host halo, $t_{\rm orb} \propto \tau_{\rm dyn} \propto [\Delta(z)\rho_{\rm crit}(z)]^{-1/2}$. There is significant scatter, however, in the apocenter, $r_{\rm apo}/R_{\rm vir}$, which results in a comparable scatter in $t_{\rm orb}/\tau_{\rm dyn}$ (Figure \ref{fig:tovertau}). The scatter originates in the concentration of the infalling subhalos: low-concentration subhalos begin to be disrupted earlier and thus experience much longer initial orbits than subhalos with higher concentrations.

\vspace{\bulletskip}
{\bf Spherical Symmetry:}
Motion in a spherical potential, in which the direction of angular momentum does not change, is not a good approximation to subhalo orbital dynamics. In particular, the direction of the angular momentum vector is not fixed. After a few orbits, the direction of the angular momentum is essentially randomized. This seems to be a generic feature of subhalo evolution in highly inhomogeneous, triaxial host halos (Figure \ref{fig:angmom}).

\vspace{\bulletskip}
{\bf Evolution in the $v_{\rm max}$--$r_{\rm max}$ plane:} As subhalos are tidally disrupted by the host halo, their maximum circular velocities and radii steadily decrease, tracing out tracks in the $v_{\rm max}$--$r_{\rm max}$ plane. While the joint median evolution, as well as their individual dependence on tidal mass, are in qualitative agreement with previous studies, we find a substantial amount of scatter. Furthermore, this scatter can be mostly attributed to variations in the concentration at infall. The difference is most pronounced in the evolution of $r_{\rm max}$: subhalos that are {\em more concentrated} at infall experience a {\em weaker evolution} in $r_{\rm max}$ as they lose mass (Figure \ref{fig:rmaxev}). Concentration at infall determines evolution in the $v_{\rm max}$--$r_{\rm max}$ plane (Figure \ref{fig:rvmaxev}).

\vspace{\bulletskip}
{\bf Disrupted Subhalo Population}: We find that subhalos with early infall and/or low concentration are preferentially disrupted within the host (Figure \ref{fig:fdisrpt}). We find no dependence of survivability on mass within the range probed by VL2. These results are consistent with the notion that tidal mass loss is correlated with concentration instead of mass for low-mass subhalos. The typical infalling mass of $m \sim 10^7\ M_\odot$ ($\mu_0 \sim 10^{-5}$) is true for both surviving and disrupted subhalos. Circularity does not influence subhalo survivability. There is a slight bias in tightly bound orbits with $\eta < 1$ being preferentially disrupted though this is a relatively small effect. The infall distributions and fitting functions for $\eta$ and $\epsilon$ presented in \S\ref{sec:results-etainf} and \S\ref{sec:results-epsinf} based on surviving VL2 subhalos should do a good job at representing the infall distribution for {\em all} subhalos (surviving plus disrupted) that ever fell onto the host (Figures \ref{fig:eta-D2} and \ref{fig:circ-D2}). This is the case at both $z_{\rm infall}$ with an NFW host potential as well as $z_{\rm cross}$ with a point mass host potential.

\vspace{\bulletskip}
In this work we have extended previous detailed analyses of subhalo dynamics and evolution to the much lower mass ratios probed by the Via Lactea II data. We have found qualitatively different behaviour in this low-mass regime, with dynamical friction and orbital dynamics playing a lesser role, and the interior structure of the subhalos,  expressed in terms of concentration, playing a much more important role.

The `concentration bias' we find here raises the prospects of significantly improving our ability to connect ultra-faint dwarf galaxies to the primordial fluctuations from which they collapsed. More detailed study, in particular with finer time resolution and a larger sample of simulated Galactic host halos, will be necessary before we can reliably use concentration bias in near field cosmology.

\acknowledgments{We are grateful to P. Madau, C. Park, and J. Taylor for useful discussions and are indebted to F. van den Bosch, M. Boylan-Kolchin, and J. Taylor for a careful reading of an earlier draft. We thank the Via Lactea collaboration (J. Diemand, M. Kuhlen \& P. Madau) for making the halo catalogs publicly available and also thank J. Diemand for clarifying some aspects of the subhalo mass assignment scheme. We acknowledge the thorough and constructive report from the anonymous referee that improved many aspects of this work. JDE acknowledges the support of the National Science and Engineering Research Council of Canada.}


\end{document}